\documentclass[]{aastex63}
\pdfoutput=1
\usepackage{lineno}
\usepackage{amsmath}

\received{June 1, 2019}
\revised{January 10, 2019}
\accepted{\today}
\submitjournal{ApJ}

\shorttitle{Interplanetary flux-rope observed by HAWC}
\shortauthors{HAWC collaboration}

\begin{document}

\title{Interplanetary magnetic flux rope observed at ground level by HAWC}

\correspondingauthor{K. P. Arunbabu}
\email{arun@igeofisica.unam.mx}
\correspondingauthor{A. Lara}
\email{alara@igeofisica.unam.mx}

\author{S.~Akiyama}
\affiliation{The Catholic University of America, Washington, DC, USA}

\author{R.~Alfaro}
\affiliation{Instituto de F\'{i}sica, Universidad Nacional Aut\'{o}noma de M\'{e}xico, Ciudad de M\'{e}xico, M\'{e}xico}

\author{C.~Alvarez} 
\affiliation{Universidad Aut\'{o}noma de Chiapas, Tuxtla Guti\'errez, Chiapas, M\'{e}xico}

\author{J.R.~Angeles Camacho}
\affiliation{Instituto de F\'{i}sica, Universidad Nacional Aut\'{o}noma de M\'{e}xico, Ciudad de M\'{e}xico, M\'{e}xico}

\author{J.C.~Arteaga-Vel\'{a}zquez}
\affiliation{Universidad Michoacana de San Nicol\'{a}s de Hidalgo, Morelia, M\'{e}xico}

\author{K.P.~Arunbabu}
\affiliation{Instituto de Geof\'{i}sica, Universidad Nacional Aut\'{o}noma de M\'{e}xico, Ciudad de M\'{e}xico, M\'exico} 

\author{D.~Avila Rojas}
\affiliation{Instituto de F\'{i}sica, Universidad Nacional Aut\'{o}noma de M\'{e}xico, Ciudad de M\'{e}xico, M\'{e}xico}

\author{H.A.~Ayala Solares}
\affiliation{Department of Physics, Pennsylvania State University, University Park, PA, USA}

\author{E.~Belmont-Moreno}
\affiliation{Instituto de F\'{i}sica, Universidad Nacional Aut\'{o}noma de M\'{e}xico, Ciudad de M\'{e}xico, M\'{e}xico}

\author{K.S.~Caballero-Mora}
\affiliation{Universidad Aut\'{o}noma de Chiapas, Tuxtla Guti\'errez, Chiapas, M\'{e}xico}

\author{T.~Capistr\'{a}n}
\affiliation{Instituto Nacional de Astrof\'{i}sica, \'{O}ptica y Electr\'{o}nica, Puebla, M\'{e}xico}

\author{A.~Carrami\~{n}ana}
\affiliation{Instituto Nacional de Astrof\'{i}sica, \'{O}ptica y Electr\'{o}nica, Puebla, M\'{e}xico}

\author{S.~Casanova}
\affiliation{Institute of Nuclear Physics Polish Academy of Sciences, PL-31342 IFJ-PAN, Krakow, Poland}

\author{P.~Colin-Farias}
\affiliation{Posgrado en Ciencias de la Tierra, Universidad Nacional Aut\'{o}noma de M\'{e}xico, Ciudad de M\'{e}xico, M\'exico}
\affiliation{Instituto de Geof\'{i}sica, Universidad Nacional Aut\'{o}noma de M\'{e}xico, Ciudad de M\'{e}xico, M\'exico} 

\author{U.~Cotti}
\affiliation{Universidad Michoacana de San Nicol\'{a}s de Hidalgo, Morelia, M\'{e}xico}

\author{J.~Cotzomi}
\affiliation{Facultad de Ciencias F\'{i}sico Matem\'{a}ticas, Benem\'{e}rita Universidad Aut\'{o}noma de Puebla, Puebla, M\'{e}xico}

\author{E.~De la Fuente}
\affiliation{Departamento de F\'{i}sica, Centro Universitario de Ciencias Exactas e Ingenier\'{i}as, Universidad de Guadalajara, Guadalajara, M\'{e}xico}

\author{C.~de Le\'{o}n}
\affiliation{Universidad Michoacana de San Nicol\'{a}s de Hidalgo, Morelia, M\'{e}xico}

\author{R.~Diaz Hernandez}
\affiliation{Instituto Nacional de Astrof\'{i}sica, \'{O}ptica y Electr\'{o}nica, Puebla, M\'{e}xico}

\author{C.~Espinoza}
\affiliation{Instituto de F\'{i}sica, Universidad Nacional Aut\'{o}noma de M\'{e}xico, Ciudad de M\'{e}xico, M\'{e}xico}

\author{N.~Fraija}
\affiliation{Instituto de Astronom\'{i}a, Universidad Nacional Aut\'{o}noma de M\'{e}xico, Ciudad de M\'{e}xico, M\'{e}xico}

\author{A.~Galv\'{a}n-G\'{a}mez}
\affiliation{Instituto de Astronom\'{i}a, Universidad Nacional Aut\'{o}noma de M\'{e}xico, Ciudad de M\'{e}xico, M\'{e}xico}

\author{D.~Garcia}
\affiliation{Instituto de F\'{i}sica, Universidad Nacional Aut\'{o}noma de M\'{e}xico, Ciudad de M\'{e}xico, M\'{e}xico}

\author{J.A.~Garc\'{i}a-Gonz\'{a}lez}
\affiliation{Instituto de F\'{i}sica, Universidad Nacional Aut\'{o}noma de M\'{e}xico, Ciudad de M\'{e}xico, M\'{e}xico}

\author{F.~Garfias}
\affiliation{Instituto de Astronom\'{i}a, Universidad Nacional Aut\'{o}noma de M\'{e}xico, Ciudad de M\'{e}xico, M\'{e}xico}

\author{M.M.~Gonz\'{a}lez}
\affiliation{Instituto de Astronom\'{i}a, Universidad Nacional Aut\'{o}noma de M\'{e}xico, Ciudad de M\'{e}xico, M\'{e}xico}

\author{J.A.~Goodman}
\affiliation{Department of Physics, University of Maryland, College Park, MD, USA}

\author{J.P.~Harding}
\affiliation{Physics Division, Los Alamos National Laboratory, Los Alamos, NM, USA}

\author{B.~Hona}
\affiliation{Department of Physics, Michigan Technological University, Houghton, MI, USA}

\author{D.~Huang}
\affiliation{Department of Physics, Michigan Technological University, Houghton, MI, USA}

\author{F.~Hueyotl-Zahuantitla}
\affiliation{Universidad Aut\'{o}noma de Chiapas, Tuxtla Guti\'errez, Chiapas, M\'{e}xico}

\author{P.~H\"{u}ntemeyer}
\affiliation{Department of Physics, Michigan Technological University, Houghton, MI, USA}

\author{A.~Iriarte}
\affiliation{Instituto de Astronom\'{i}a, Universidad Nacional Aut\'{o}noma de M\'{e}xico, Ciudad de M\'{e}xico, M\'{e}xico}

\author{V.~Joshi}
\affiliation{Erlangen Centre for Astroparticle Physics, Friedrich-Alexander-Universit\"{a}t Erlangen-N\"{u}rnberg, Erlangen, Germany}

\author{D.~Kieda}
\affiliation{Department of Physics and Astronomy, University of Utah, Salt Lake City, UT, USA}

\author{G.J.~Kunde}
\affiliation{Physics Division, Los Alamos National Laboratory, Los Alamos, NM, USA}

\author{A.~Lara}
\affiliation{Instituto de Geof\'{i}sica, Universidad Nacional Aut\'{o}noma de M\'{e}xico, Ciudad de M\'{e}xico, M\'exico} 

\author{H.~Le\'on Vargas}
\affiliation{Instituto de F\'{i}sica, Universidad Nacional Aut\'{o}noma de M\'{e}xico, Ciudad de M\'{e}xico, M\'{e}xico} 

\author{G.~Luis-Raya}
\affiliation{Universidad Polit\'{e}cnica de Pachuca, Pachuca, Hidalgo, M\'{e}xico}

\author{K.~Malone}
\affiliation{Physics Division, Los Alamos National Laboratory, Los Alamos, NM, USA}

\author{J.~Mart\'inez-Castro}
\affiliation{Centro de Investigaci\'on en Computaci\'on, Instituto Polit\'{e}cnico Nacional, Ciudad de M\'{e}xico, M\'{e}xico}

\author{J.A.~Matthews}
\affiliation{Dept of Physics and Astronomy, University of New Mexico, Albuquerque, NM, USA}

\author{P.~Miranda-Romagnoli}
\affiliation{Universidad Aut\'{o}noma del Estado de Hidalgo, Pachuca, M\'{e}xico}

\author{E.~Moreno}
\affiliation{Facultad de Ciencias F\'{i}sico Matem\'{a}ticas, Benem\'{e}rita Universidad Aut\'{o}noma de Puebla, Puebla, M\'{e}xico}

\author{A.~Nayerhoda}
\affiliation{Institute of Nuclear Physics Polish Academy of Sciences, PL-31342 IFJ-PAN, Krakow, Poland}

\author{L.~Nellen}
\affiliation{Instituto de Ciencias Nucleares, Universidad Nacional Aut\'{o}noma de M\'{e}xico, Ciudad de M\'{e}xico, M\'{e}xico}

\author{M.~Newbold}
\affiliation{Department of Physics and Astronomy, University of Utah, Salt Lake City, UT, USA}

\author{T.~Niembro}
\affiliation{Smithsonian Astrophysical Observatory, Cambridge, MA, USA}

\author{T.~Nieves-Chinchilla}
\affiliation{Heliosphysics science division, NASA- GSFC, Washington DC, USA}

\author{R.~Noriega-Papaqui}
\affiliation{Universidad Aut\'{o}noma del Estado de Hidalgo, Pachuca, M\'{e}xico}

\author{E.G.~P\'erez-P\'erez}
\affiliation{Universidad Polit\'{e}cnica de Pachuca, Pachuca, Hidalgo, M\'{e}xico}

\author{L.~Preisser}
\affiliation{Instituto de Geof\'{i}sica, Universidad Nacional Aut\'{o}noma de M\'{e}xico, Ciudad de M\'{e}xico, M\'exico} 

\author{C.D.~Rho}
\affiliation{University of Seoul, Seoul, Rep. of Korea}

\author{J.~Ryan}
\affiliation{University of New Hampshire, NH, USA}

\author{H.~Salazar}
\affiliation{Facultad de Ciencias F\'{i}sico Matem\'{a}ticas, Benem\'{e}rita Universidad Aut\'{o}noma de Puebla, Puebla, M\'{e}xico}

\author{F.~Salesa Greus}
\affiliation{Institute of Nuclear Physics Polish Academy of Sciences, PL-31342 IFJ-PAN, Krakow, Poland}

\author{A.~Sandoval}
\affiliation{Instituto de F\'{i}sica, Universidad Nacional Aut\'{o}noma de M\'{e}xico, Ciudad de M\'{e}xico, M\'{e}xico}

\author{R.W.~Springer}
\affiliation{Department of Physics and Astronomy, University of Utah, Salt Lake City, UT, USA}

\author{I.~Torres} 
\affiliation{Instituto Nacional de Astrof\'{i}sica, \'{O}ptica y Electr\'{o}nica, Puebla, M\'{e}xico}

\author{F.~Ure\~{n}a-Mena}
\affiliation{Instituto Nacional de Astrof\'{i}sica, \'{O}ptica y Electr\'{o}nica, Puebla, M\'{e}xico}

\author{L.~Villase\~{n}or}
\affiliation{Facultad de Ciencias F\'{i}sico Matem\'{a}ticas, Benem\'{e}rita Universidad Aut\'{o}noma de Puebla, Puebla, M\'{e}xico}

\author{A.~Zepeda}
\affiliation{Departamento de F\'{i}sica, Centro de Investigaci\'{o}n y de Estudios Avanzados del Instituto Polit\'{e}cnico Nacional, Ciudad de M\'{e}xico, M\'{e}xico}

\begin{abstract}

We report the ground-level detection of a Galactic cosmic-ray (GCR) flux enhancement lasting ~$\sim$~17~hr and associated with the passage of a magnetic flux rope (MFR) over the Earth. The MFR was associated with a slow Coronal Mass Ejection (CME) caused by the eruption of a filament on 2016 October 9. Due to the quiet conditions during the eruption and the lack of interactions during the interplanetary~CME transport to the Earth, the associated MFR preserved its configuration and reached the Earth with a strong magnetic field, low density, and a very low turbulence level compared to local background, thus generating the ideal conditions to  redirect and guide GCRs (in the  $\sim$ 8 to 60 GV rigidity range) along the magnetic field of the MFR.  
An important negative $B_Z$ component inside the MFR caused large disturbances in the geomagnetic field and a relatively strong geomagnetic storm. However, these disturbances are not the main factors behind the GCR enhancement. Instead, we found that the major factor was the alignment between of the MFR axis and the asymptotic direction of the observer.

\end{abstract}


\section{Introduction}
\label{sec:intro}

The modulation of the Galactic cosmic-ray (GCR) flux by solar activity has been known and studied since the 1930s, after the discovery of these extraterrestrial  particles \citep[see][for a historical review of this subject]{Lockwood1971}. The Parker theory of transport \citep{Parker1965} has been  successfully applied to explain the ``long-term modulation," i.e., the observed anticorrelation between the sunspot number and the GCR flux on the time scale of the solar magnetic cycle \citep[22 yr,][]{Ferreira_2004}. Moreover, analyzing direct measurements of the interplanetary magnetic field over an entire solar cycle \citet{1983JGR....88.2973D} reached the conclusion that GCR intensity increases and decreases where associated with weak and strong interplanetary magnetic fields, respectively. 

On shorter time scales (weeks or days) the modulation is driven by disturbances such as coronal mass ejections (CMEs) which are large amounts of magnetized plasma \citep[$\sim 10^{15}$ g][]{Colaninno_2009} expelled from the low corona to the interplanetary medium with speeds ranging from $\sim 100$ to $\sim 3000$ km s$^{-1}$. At low coronal levels, the major events associated with CME are flares \citep{2015SoPh..290.3457S} and filament eruptions \citep{2019ApJ...880...84S}, the latter, also known as prominences constitute the tracers of helical flux ropes in the corona \citep{2015JApA...36..157F}.

The Interplanetary counterparts of CMEs (ICMEs) can be recognized by their structure: a leading shock wave, followed by a turbulent sheath region and the driving ejecta \citep{2020SoPh..295...61L}. A particular subset of ICMEs \citep[up to 77\% according to][]{2019SoPh..294...89N} shows a clear rotation of the magnetic field components corresponding to a magnetic flux rope (MFR) structure known as magnetic cloud \citep[MC, ][]{1981JGR....86.6673B}. From the space weather point of view, ICMEs have deserved great attention due to their potential effects on the Earth's magnetic field and the associated risk to the space-based technology. These effects are known as ``geo-effectiveness'' and are generally measured by geomagnetic indices such as the so-called Disturbance Storm-Time (D$_{ST}$) index \footnote{``The D$_{ST}$ index is related to the average of the longitudinal component of the magnetic external field measured at the equator and surface of the Earth assumed as a dipole'' \citep{osti_4554034}. We recommend the reading of \citet{2017SSRv..212.1159M,2017LRSP...14....5K,2020SoPh..295...61L} and all references therein for further information about the connection between Magnetic Flux-Rope (MFR), CME, ICME and MC.}. 

It is widely accepted that ICMEs may cause decrements of the GCR flux known as Forbush decreases \citep[FD,][]{1937PhRv...51.1108F}. 
Extensive literature has been devoted to the detailed study of the effects of ICME  internal structure on the GCR intensity \citep[e.g.][and references therein]{1963JGR....68.6361S,2019SoPh..294...86S,2020ApJ...896..133L}. In particular, there is no consensus about the role that the shock, the sheath, and the MC plays on the modulation \citep{Cane2000, 2011SoPh..270..609R}. Furthermore, the not-uncommon magnetic field topology of a helical flux tube or MFR has been poorly studied, particularly in terms of the possibility of how the force-free and helical geometry affects the local GCR population by redirecting and guiding of GCRs if the turbulence intensity is low  \citep{Belov2015}.

On time scales of hours,
GCR enhancements have been detected before the arrival of CMEs; these are the so-called precursors and are attributed to a loss-cone mechanism \citep{2000JGR...10527457M,2011GeoRL..3816108R}.
Besides,
enhancements of the GCR intensity have been observed during geomagnetic storms since the 1950’s \citep{1959Natur.183..381Y}. Soon thereafter, it was noticed that these increases were not caused by solar energetic particles (related to flares), nor by anomalies of the diurnal anisotropy; neither were they due to changes in the cutoff rigidity caused by the variations of the geomagnetic field \citep{Kondo1960}. 
\citet[][]{1963ICRC....3..421A} outlined the possibility that a ``magnetic tongue" rooted in the Sun was causing the observed GCR enhancements 
\citep[see][for more proposed mechanisms]{1962PhRvL...8..215D}. Nevertheless, the poor knowledge of the ICME structure at that time prevented the advance of the investigations and the subject was somehow forgotten.

An important GCR enhancement was observed in 2016  October by the High Altitude Water Cherenkov array (HAWC), and in this work we present evidence that this excess was due to an anisotropic GCR flux caused by the MFR.
The size and the energy range of HAWC (Section~\ref{sec:hawc}) enabled the detection of this enhancement, which was  less than $\sim$~1\% with respect to the mean GCR flux for rigidities $> 8$ GV, a marginal signal at best for the neutron monitor (NM) network.

Besides the high sensitivity of HAWC, we propose that three major factors contributed to this unprecedented observation. Therefore we explain each of those factors in detail, namely:
\begin{itemize}
\item 
The solar origin of the MFR was a ``quiet'' filament eruption (i.e. without an associated flare) resulting in a slow CME ($\sim$ 180 km $s^{-1}$) launched on 2016 October 9 \citep[Section~\ref{sec:sun} and see][]{2017SoPh..292..125N}. 

\item 
The MFR reached 1 AU without being seriously distorted or disrupted, preserving a regular geometry and low turbulence level. Due to the lack of observations between the Sun and 1 AU, we performed 2D hydrodynamic simulations of the interplanetary transport of the MFR (Section~\ref{sec:sim}) and found that the associated ICME transport was surrounded, front and back, by two high-speed streamers (HSS, \citet{1963JGR....68.6361S, 1976SoPh...49..271S}), but without interacting with them, preserving helical geometry properties of the the MFR  at 1~AU (Section~\ref{sec:interplanetary}).

\item
The impact of the MFR on the Earth's magnetosphere was observed by a constellation of near-Earth spacecraft. The low flow pressure combined with the large south $B_Z$ component of the field in the MFR, produced intense and moderate  disturbances of the magnetosphere the day before and during the HAWC observations respectively (Section~\ref{sec:magnetosphere}). This lack of correlation shows that the geomagnetic disturbances are not the main causes of the HAWC observation.

\item The main factor that allowed detection at ground-level is the alignment between the MFR axis and the asymptotic direction of the detector (Section \ref{sec:model}).

\end{itemize}

\begin{figure}[ht!]
\includegraphics[width=0.6\columnwidth]{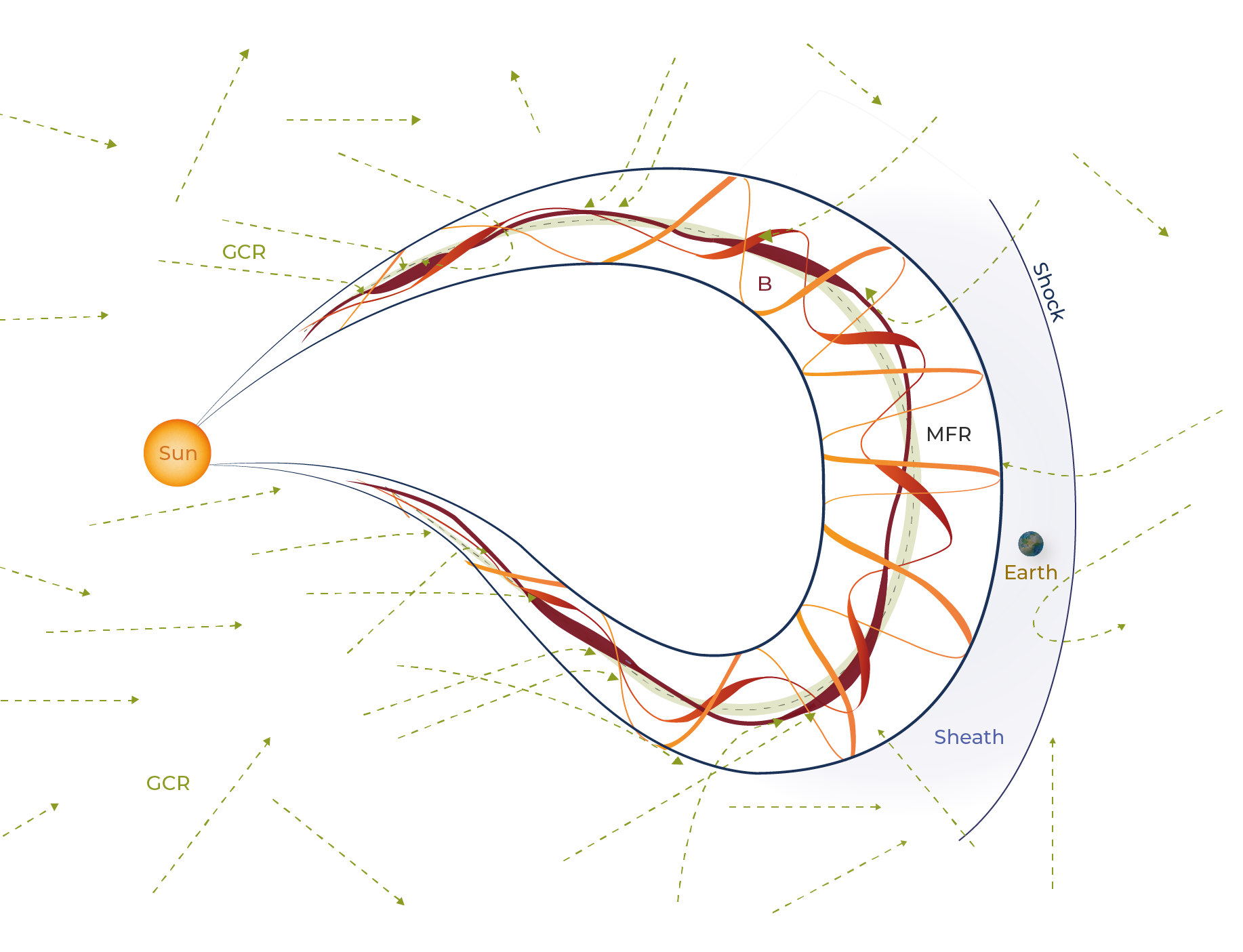}\centering
\caption{Schematic representation of an interplanetary MFR (the field magnitudes are represented by different reddish color bands,  the darker and thicker bands corresponding to a stronger magnetic field) arriving at the vicinity of the Earth. The GCRs (dashed green lines) are deflected by the helical magnetic field and guided along the MFR axis.}
\label{fig:cartoon}
\end{figure}

With the absence of significant solar activity, a marginal registration of the event in the NM network, and no other corroborating high-energy signal, one would normally relegate the HAWC signal to a category of unexplained features. However, due to the confluence of several heliospheric conditions, we suggest a model in which the GCRs are guided along the MFR axis (a schematic of this scenario is shown in Figure~\ref{fig:cartoon}), which can explain the phenomenon. As stated, the model relies on these several conditions and we justify each of them via observation or simulation, the key ones being listed above.

To support our findings, we have fitted an MFR model to the {\it in-situ} observations using the circular-cylindrical model \citep{2016ApJ...823...27N}. Then, with the fitted MFR parameters, we computed the trajectory of the GCRs inside the MFR and found a good match (in time and energy) between the trajectories of the GCRs guided inside the MFR and the GCR enhancements observed by HAWC (Section~\ref{sec:fr_gcr}). Finally, the discussion and our conclusions are presented in Sections~\ref{sec:discussion} and \ref{sec:conclusion}. 

\section{HAWC and its observation  of a Flux Rope }  
\label{sec:hawc}

\subsection{HAWC observatory} 
The High Altitude Water Cherenkov (HAWC) observatory is located on a plateau between the Sierra Negra and Pico de Orizaba Volcanoes, at $18^{\circ}59' 41"$ N, $97^{\circ} 18'30.6"$ W and at an altitude of 4100 m. HAWC consists of 300 water Cherenkov detectors (WCD), each one 7.3 m in diameter and 4.5 m deep. The WCDs are spread over an area of 22,000 $m^2$ and each WCD is filled with filtered water and instrumented with 4 Photo-Multiplier Tubes (PMTs). A 10-inch PMT is located at the center of the WCD and three 8-inch PMTs are arranged around the central one making an equilateral triangle of side 3.2 m  \citep{Abeysekara_HAWC2012}.

The main DAQ system measures arrival times and time over threshold of PMT pulses. This information is used to reconstruct the air shower front and estimate the arrival direction and energy of the primary particles. The electronics are based on Time-to-Digital Converters (TDC), and the main DAQ also has a scaler system which counts the hits inside a time window of 30 ns for each PMT ($R_1$ rate from now on) and the coincidences of 2, 3 and 4 PMTs in each WCD, called multiplicity $M_2$, $M_3$ and $M_4$, respectively. The TDC-scaler system allows one to measure particles with relatively low rigidity, from the cutoff rigidity at HAWC site ($\sim 8$ GV) to the low limit of the reconstructed showers ($\sim 100$ GV). The median energy of observation of TDC-scaler system and multiplicities are in the range of 40-46 GV. The large area and low cutoff rigidity of HAWC make it a suitable instrument for studying solar modulation in general and space weather in particular. Summarizing, we can say that the TDC-scaler rate used in this work provides information on the primary GCR flux above cutoff rigidity (from 8 GV onwards) reaching Earth's atmosphere which can be measured with high precision. 

\subsection{TDC-scaler data reduction}

During long periods of observation, the efficiency of PMTs may vary, so to carry out high precision studies, it is necessary to correct for these drifts. To perform these corrections, first we have identified relatively stable PMTs during a year by continuously checking their deviations, and using a ``singular value decomposition" method, we compute a reference rate which was used to model the slow and small changes in the efficiency of the remaining PMTs and correct for their efficiency variation. 
Due to the large collecting area and high altitude of HAWC, the rate of observed particles is high, as instance during  the year 2016 the efficiency corrected average rate of the HAWC TDC-scaler system was  $<R_{1}>$ = 23.39 kHz per PMT, similarly for multiplicities the average rates per WCD were $<{M_2}>$ = 8.06 kHz, $<{M_3}>$ = 5.69 kHz, and $<{M_4}>$ = 4.35 kHz.  The analysis carried out in this study is using data with `one minute' resolution, and for uniformity the rates were converted to percentage deviation with respect to the mean rate of the year 2016. It is worth mentioning that after the efficiency  correction, the measured TDC-scaler rates are in agreement with the expected statistical accuracy of the data. 
Similar to other air shower detectors, the TDC-scaler rates show a dependence on barometric pressure. We correct this pressure modulation following the method described by \citet{Arun2019}. 
Finally, due to its near equator location and lower cutoff rigidity of 8 GV of HAWC, the solar diurnal anisotropy (DA) component is strongly significant in its low energy observations. The DA was removed from the data using a band rejection filter that removes all the frequencies within the frequency range of $\sim 1 - 2$ cycles per day. 

The efficiency corrected TDC-scaler rates of all the PMTs were combined to provide the $R_1$ rate with a statistical accuracy $<0.01\%$ per minute. The same efficiency correction process was also applied to the multiplicity rates from all the WCDs and these were combined to get $M_2$, $M_3$ and $M_4$. 

\subsection{TDC-Scaler and Solar Modulation} 

\begin{figure}[htbp]
\includegraphics[width=\textwidth]{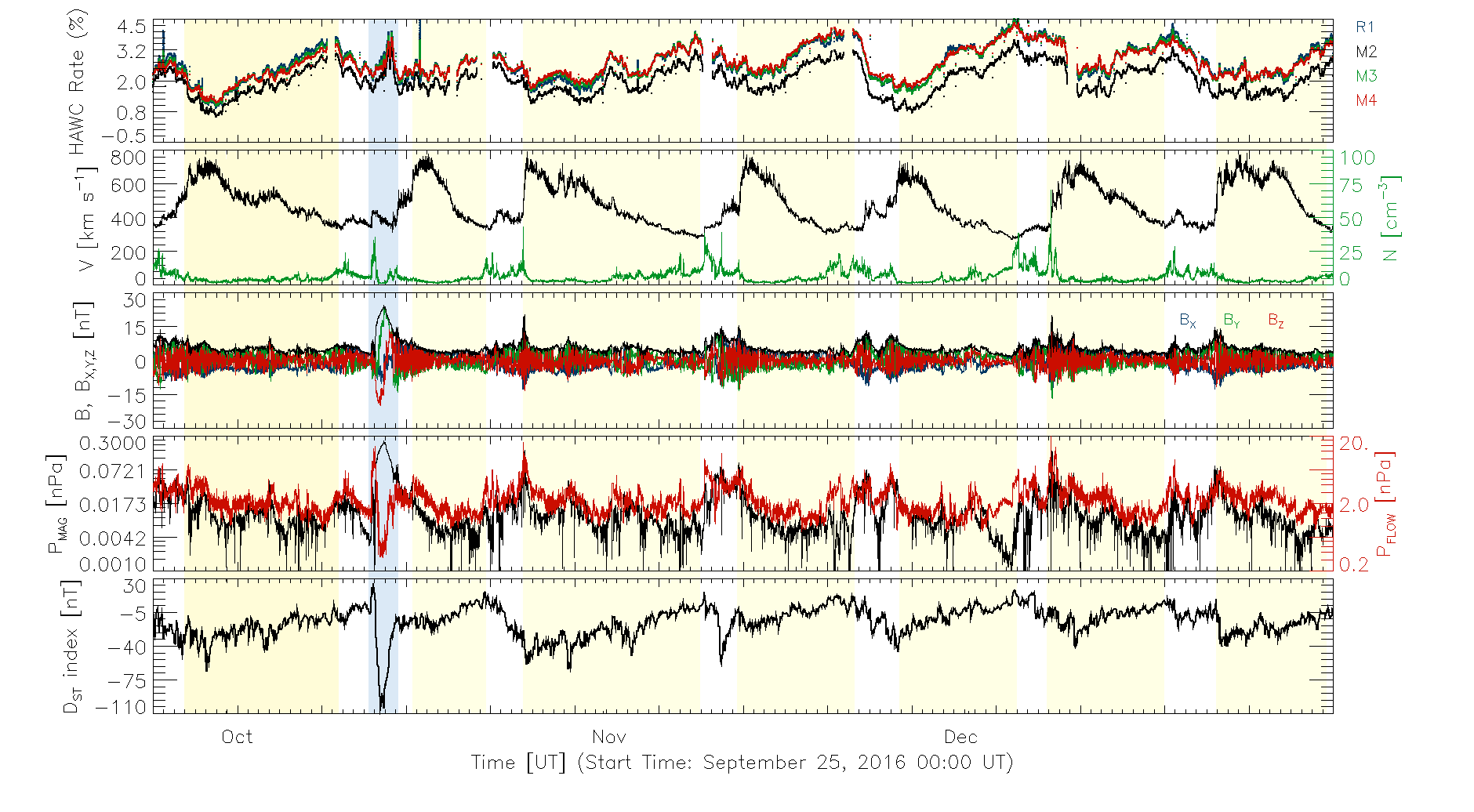}
\caption{From top to bottom, we show:  HAWC TDC-scaler rates ($R_1$, $M_2$, $M_3$ and $M_4$, blue, black, green and red, respectively); from the OMNI data, the SW measurements of: the speed V and the number density N (both shown in second panel in black and green solid lines); the components (B$_X$, B$_Y$ and, B$_Z$) and magnitude of the $B-$field (shown in third panel and colored in blue, green, red and black, respectively); the magnetic and flow pressures (black and red lines in fourth panel), and the D$_{ST}$ index (bottom panel) from the World Data Center for Geomagnetism, Kyoto for the period of time between 2016 September 25 to 2016 December 31. The yellow shading areas show the periods of time within HSSs, whereas the blue shading marks the MFR time. The narrow peaks seen in the TDC-scaler rates on October 1 and 16 are due to atmospheric electric field disturbances.} 
\label{fig:interplanetary}
\end{figure}

\begin{figure}[htbp]
\includegraphics[width=1.0\textwidth]{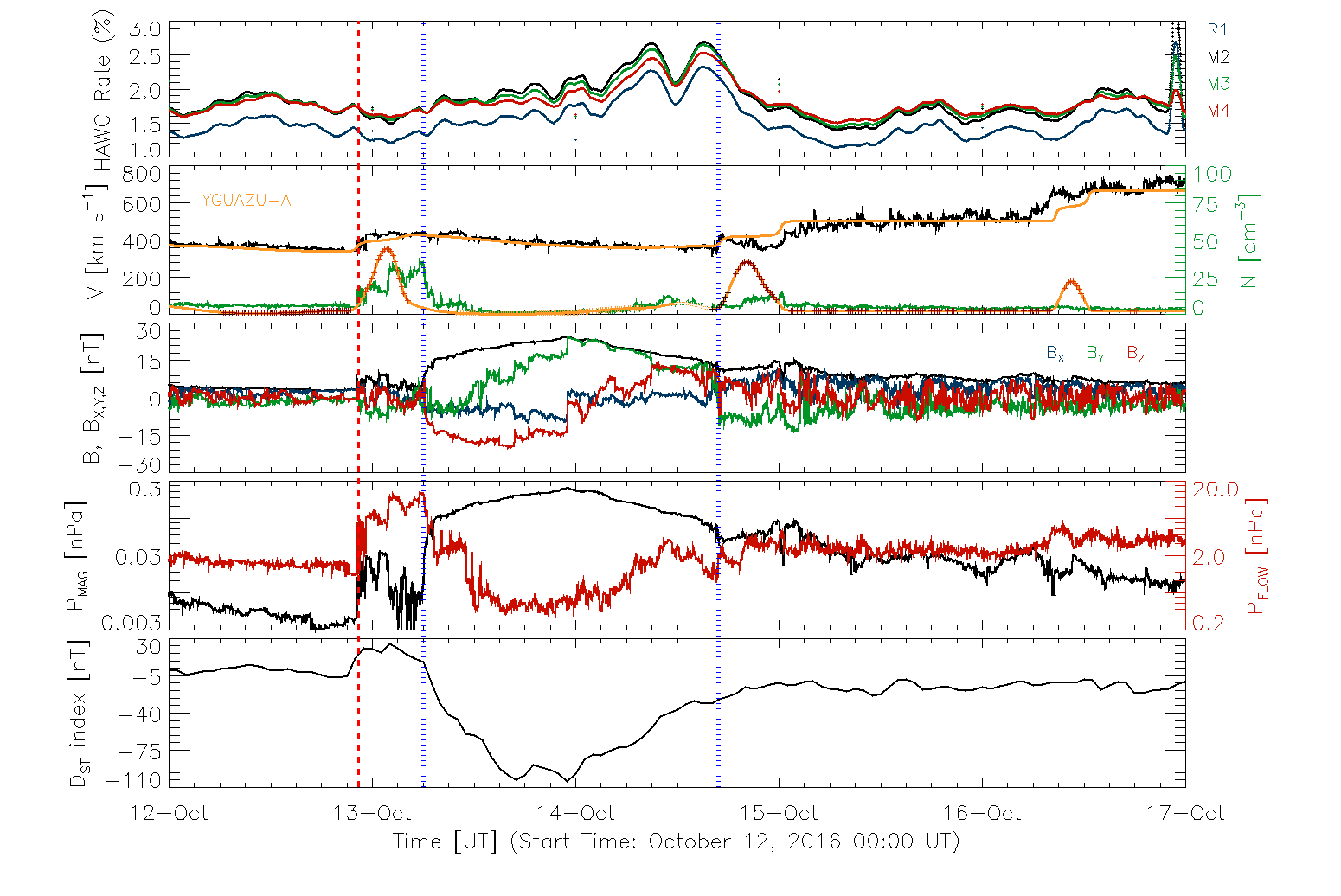}\centering
\caption{The MFR observed at the interplanetary medium, the associated geomagnetic storm and its effects over the GCR flux. From top to bottom: GCR rates measured by  HAWC TDC-scaler system (the PMT and multiplicity rates are marked with different colors, see Section \ref{sec:hawc} for their definitions); speed V (show in black) and number density N (green); the components (B$_X$, B$_Y$ and, B$_Z$, colored in blue, green and red, respectively) and magnitude of the magnetic field B (black); the magnetic (black) and flow (red) pressures; and the D$_{ST}$ (Disturbance Storm-Time) index (black). The SW parameters are extracted from the OMNI data \citep{2005JGRA..110.2104K} while the D$_{ST}$ from the World Data Center for Geomagnetism, Kyoto \citep{https://doi.org/10.17593/14515-74000}.. In the second panel, we overlapped in orange, the N and V synthetic profiles obtained with YGUAZ\'U-A code (the simulation is described in section \ref{sec:sim}). The code enables us to tag and follow different plasma parcels marked in the synthetic number density profile with colored symbols as described in Table~\ref{tabla:sw}. The dashed vertical red line marks the shock and the two dotted blue lines mark the start and end time of the slow CME constraining a MFR. The TDC-scaler rate increase observed at October 16 at $\sim$ 23 UT was caused by an atmospheric electric field transient and is unrelated to this investigation \citep{2017ICRC...35...80L,2019ICRC...36.1087J}.}
\label{fig:fluxropeSW}
\end{figure}

The TDC-scaler data after applying all these corrections are shown in the top panels of Figures~ \ref{fig:interplanetary} and \ref{fig:fluxropeSW}
where the mean values of the four channels $R_1$ (blue), $M_2$ (black), $M_3$ (green), and $M_4$ (red) are shown. 
Along with these figures also shown in the second and third panels the Solar Wind (SW) proton measurements of: the speed V (black), the number density N (green); the magnetic field magnitude B (black) and its components B$_X$(blue), B$_Y$ (green) and, B$_Z$ (red). The one-minute resolution SW parameters were obtained from the OMNI data service developed and supported by NASA/Goddard's Space Physics Data Facility\footnote{The OMNI one-minute resolution data set is created from ACE, Wind and DSCOVR in-situ measurements \citep{2005JGRA..110.2104K}. These spacecraft are located at L1 point.}. The flow and magnetic pressures (fourth panel) as well as the D$_{ST}$ index (bottom panel) were obtained from the World Data Center for Geomagnetism, Kyoto, \citet{https://doi.org/10.17593/14515-74000}.

In order to show 
the suitability of the HAWC TDC-scaler system to study the solar modulation at short and medium time scales, Figure~\ref{fig:interplanetary} shows a 
time period of approximately three months at the end of 2016, where seven HSSs (marked with yellow shadow areas) and one ICME and its MFR (blue shadow) were observed. From this figure the visible correlation of HAWC rates with the SW velocity resembles the advection effect of the GCRs in the Heliosphere. Also the diffusion effects of GCRs inside the Heliosphere show correlated decreases in HAWC rate with the magnetic field enhancements. It should be noted that, contrary to expected, the signature of the MFR structure caused an increase in the rates, which is fairly remarkable (the aim of this analysis is to study the origin of this increase).
In similar way, the changes in the magnetic field, flux and magnetic pressures and D$_{ST}$ index caused by this MFR are clearly distinguished from the rest of SW structures detected over this period. 

\subsection{TDC-Scaler Rates during the MFR Passage}

The GCR modulation associated with the ICME passage during 2016 October  observed by HAWC can be seen in Figure~\ref{fig:fluxropeSW} (see Section \ref{subsec:ICME} for the details of the  Figure panels). During this time, the TDC-scaler rates increased as a double peak structure, starting around 02:50 UT on 2016 October 14 and lasting 16.8 hr.

\begin{figure}[htbp]
\includegraphics[width=1.0\columnwidth]{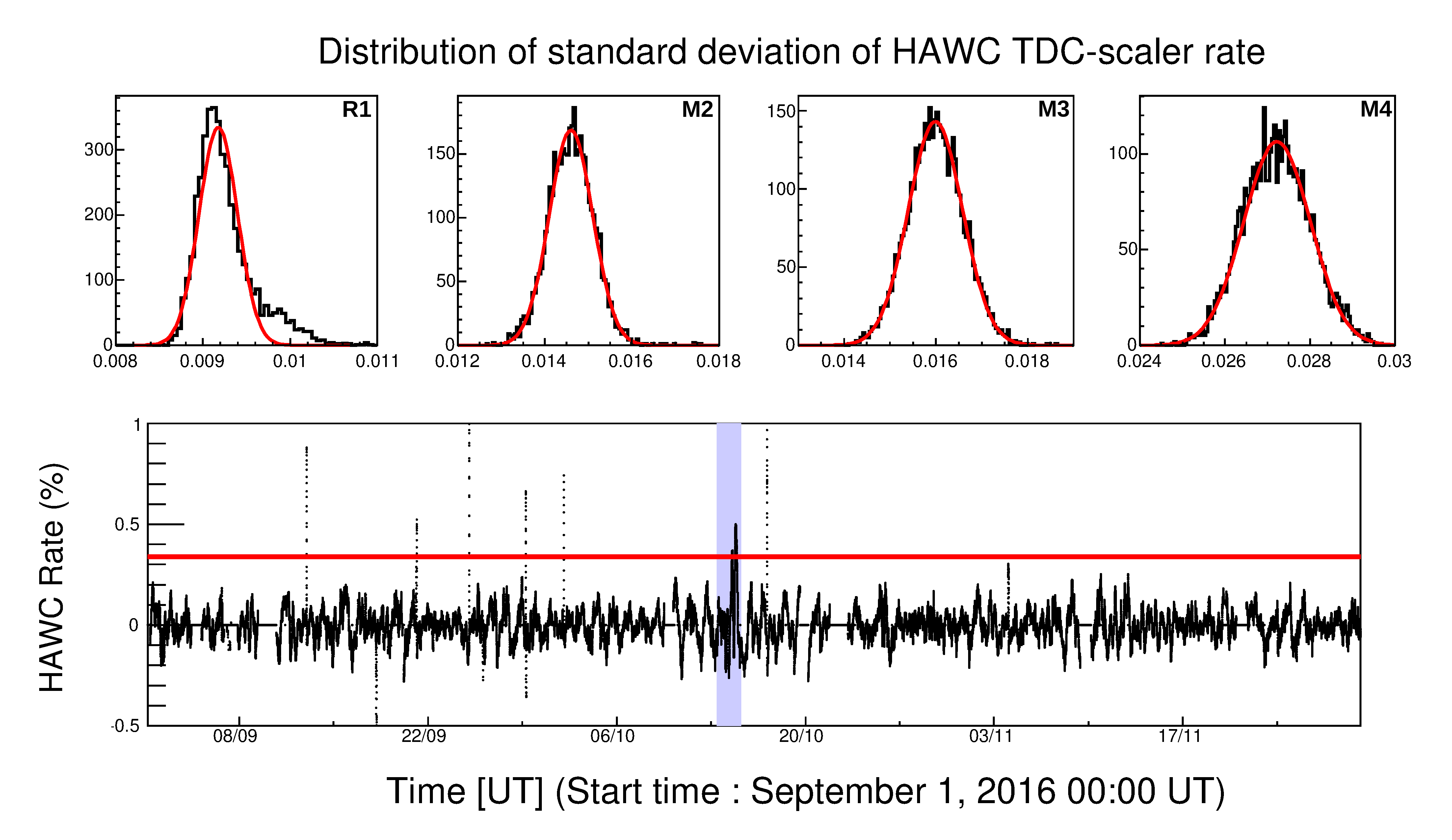}
\centering
\caption{The upper panels show the distributions of 1 minute standard deviation of the HAWC TDC-scaler rates. The distributions cover a period of three day observations: 13, 14 and 15 of October 2016. 
The bottom panel shows the HAWC TDC-scaler rate $R1$ after applying a high-pass filter, which have frequencies higher that $day^{-1}$. The red line shows the limit of 5 times standard deviation of this high-pass filtered data and the blue shaded region mark the period of time when the MFR was observed.
\label{app:Dsig}}
\end{figure}

The HAWC TDC-scaler rates have a background rate of $4.25 \times 10^8$ particles per minute. During this event, the integrated counts had an excess of $1.76 \times 10^9$ particles above the background-level.
In order to quantify the noise level of the TDC-scaler system, we computed the mean rates and  standard deviation  of the 1200 PMT rates for $R_1$ and 300 WCD rates for multiplicities $M_2$, $M_3$ and $M_4$ every minute, the upper panels of Figure \ref{app:Dsig} shows the distributions of the computed  standard deviations during three days of our period of interest. 
The mean value of these distributions are 
used as standard deviation of the observations ($\sigma$) for this study and are listed in column two of Table \ref{tabla:obsH}.
 The magnitude of the observed double peak structure in percentage, is estimated as the difference between the peak and pre-event intensities and are shown in the third and fifth columns of Table \ref{tabla:obsH}. In similar way, the magnitude in terms of standard deviation  ($\sigma$),  i. e., the Magnitude/$\sigma$, which we call as significance of the increase are  shown in fourth and sixth columns.

\begin{table*}[htbp] 
\centering
\caption{Magnitude in terms of percentage deviation (difference between the peak and pre-event intensities) of the 
first and second peaks within the double peak structure as 
observed in the HAWC TDC-scaler rate ($R_1$) and multiplicities ($M_2$, $M_3$ and $M_4$) are given in columns three and five. The standard deviation during the observation ($\sigma$) of the
TDC-scaler rate is shown in column two, and the
significance of the increases or magnitudes in terms of $\sigma$ are given in columns four and six.}
\label{tabla:obsH}
\begin{tabular}{|l|c|c|c|c|c|}  \hline
TDC-Scaler &  $\sigma$ & \multicolumn{2}{c|}{Magnitude of Peak 1 }                & \multicolumn{2}{c|}{ Magnitude of Peak 2} \\  \cline{3-6}
    &     (\%)        &  (\%) &  in terms of  $\sigma$  &  (\%) &  in terms of $\sigma$ \\ \hline  
$R_1$ & $9.18 \times 10^{-03}$ & 0.7122 & 77.6 & 0.7761  & 84.6 \\
$M_2$ & $1.46 \times 10^{-02}$ & 0.7562 & 51.8 & 0.7843  & 53.7 \\
$M_3$ & $1.60 \times 10^{-02}$ & 0.7235 & 45.2 & 0.7940  & 49.7 \\
$M_4$ & $2.72 \times 10^{-02}$ & 0.6690 & 24.6 & 0.7570  & 27.8 \\ \hline 
 \end{tabular}
\end{table*}

It should be noted that during the event, the weather at HAWC site was calm and no electric field disturbances were recorded by the site electric field monitor. The atmospheric pressure at the site was also showing normal behavior, which rules out the possibility of abnormal atmospheric activities. Earthquakes can also cause rate increases in the TDC-scaler system but these last a few minutes and there is no report of an earthquake of magnitude greater than 5.5  during our period of interest in the south-central part of Mexico, where HAWC is placed.

To compare this event with other short-term modulation observed by the TDC-scaler system, we applied a high-pass filter to the rates. This filter removed all the frequencies smaller than $day^{-1}$ and retained all the fast variation in the data that have time scales less than a day, the bottom panel of Figure~\ref{app:Dsig} shows the high-pass filtered data during three months, starting on September 1, 2016. The standard deviation of this high-pass filtered data for the entire year 2016 was estimated as $\sigma_{high-pass}$=~0.068\%. From the figure it is clear that the MFR event (marked with a blue shadow area) is standing significant in comparison to all other short period modulations, which is much higher that 5 $\sigma_{high-pass}$ level. In this figure one can also see atmospheric electric field transients (short time spikes of tens of minutes). In contrast, the event due to MFR was having a duration of $\sim 17$ hours. 
These atmospheric electric field events except on September 21 and 26 are listed in \citet{2019ICRC...36.1087J}, these two days were not listed due to short gaps in HAWC TDC-scaler data.  \\

\section{Magnetic Flux-Rope Origin, Transport and Geomagnetic Effects} 
\subsection{The quiet filament eruption}  \label{sec:sun}

\begin{figure}[htbp]
\begin{center}
\includegraphics[width=0.9\textwidth]{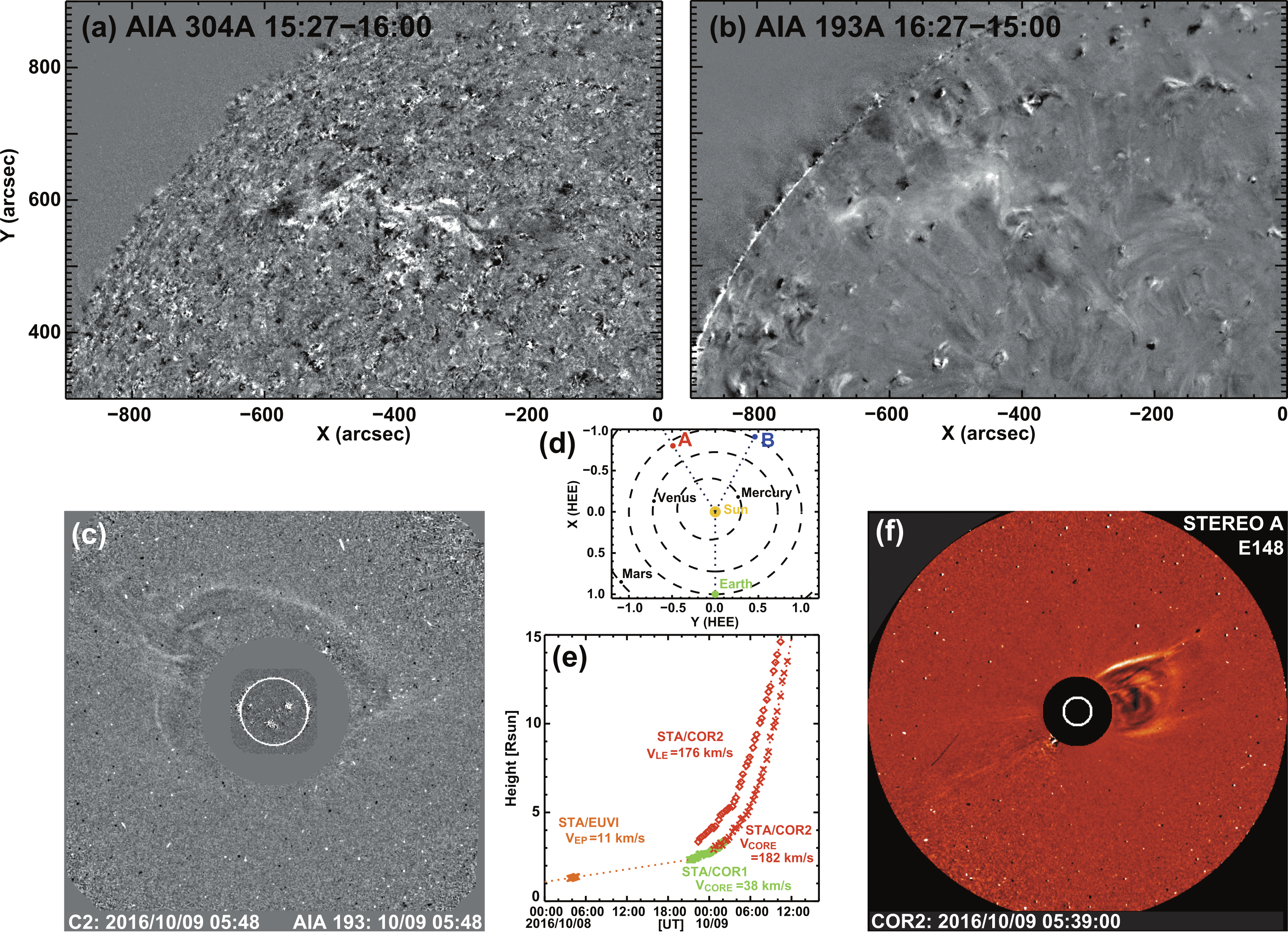}
\caption{Two stages of the October 8, 2016 Filament Eruption seen in the \textit{running differences} images of AIA 304 \r{A} at 15:27 UT and AIA 193 \r{A} at 16:27 UT (white and black continuous areas at the center of panels a and b, respectively). The associated halo and limb CME seen by LASCO (aboard SOHO spacecraft) and SECCHI (aboard STEREO-A spacecraft) are shown in panels c and f, respectively. The relative positions of the spacecrafts in the interplanetary space are shown in panel d where SOHO and Wind spacecraft are close to the Earth (marked in green) and STEREO A and B spacecraft positions are marked with red and blue colors respectively. The height-time diagram associated to the filament (orange crosses) and CME leading edge (red diamonds) and core (green and red crosses) evolution are plotted in panel e.}
\label{fig:sun}
\end{center}
\end{figure}

On 2016 October 8 a filament was observed by the Atmospheric Imaging Assembly \citep[AIA, ][]{2012SoPh..275...17L} on board of the Solar Dynamics Observatory  \citep[SDO, ][]{2012SoPh..275....3P} activated at $\sim$ 04:00 UT in the northeast quadrant (see panels a and b of Figure \ref{fig:sun}), rising up with a speed of $\sim$ 20 km s$^{-1}$ during the next 20 hours. Around  00:00 UT on October 9, 2016 and at an altitude of $\sim 3 R_\odot$ an acceleration process started as seen by COR1 and COR2, instruments on board of the Solar Terrestrial Relations Observatory \citep[STEREO,][]{2008SSRv..136....5K}  spacecraft (green and red X symbols, respectively in panel e of Figure~\ref{fig:sun}) reaching a speed of $\sim$ 180 km s$^{-1}$. 

The filament was part of a halo CME observed on October 9, 2016 at 02:45 UT by the Large Angle and Spectrometric Coronagraph (LASCO) on board of the Solar \& Heliospheric Observatory (SOHO). This eruption was seen as a limb CME, showing a dark cavity surrounded by bright material \citep[typical structure of the cross section of  MFR, e. g.,][]{1999ApJ...516..465D} by the Sun Earth Connection Coronal and Heliospheric Investigation (SECCHI) on board of STEREO-A spacecraft (panels c and f of Figure~\ref{fig:sun}, respectively). 

\subsection{The magnetic flux-rope transport in the interplanetary medium} \label{sec:interplanetary}

The ICME propagation from the Sun to the Earth of the 2016 October event has been studied previously by \citet{He2018} who combined remote sensing observations from SDO, STEREO, and SOHO and in situ measurements as extracted from OMNI data. These authors focused their work on understanding the geo-effectiveness associated with the CME magnetic field structure. They reconstructed the CME using the Grad-Shafranov technique to identify the MFR with a strong southward magnetic field, which produced the geomagnetic storm \citep[see Figure 8 of][]{He2018}. They also simulated the SW with the ENLIL-MHD code \citep{2004JGRA..109.2116O} to illustrate the background SW conditions during the ICME propagation. However, they only considered the evolution of the ICME bracketed between two HSS, without considering that all this system is propagating within a variable SW. We note that in the case of slow CMEs the Sun-Earth travel time may be a few days and therefore they have the opportunity to interact with other SW structures and/or transients that can enhance or ensure their geo-effectiveness \citep[e.g.,][]{He2018, Chen2019}.

To include this ingredient in the scenario (a variable ambient SW), we perform a simulation that includes the observed parcels of SW with different speeds and densities, including the slow ICME and the slow and fast flows (Section \ref{sec:sim}). 

\subsection{{\it In situ} observation of the ICME} \label{subsec:ICME}

When an interplanetary transient is not fast enough, it may still be able to modulate the GCR flux depending on its magnetic field strength and geometry, its dynamics compared with ambient SW, and on its interaction with the Earth's magnetic field. This means that, in general, the GCR modulation depends mainly on three factors: the nature of the transient and its intrinsic magnetic field, the resulting conditions after the transient interactats with the ambient SW, and the Earth's magnetosphere. 

During 2016 October 12-17, a slow ICME was identified at 1 AU using the one-minute resolution in-situ measurements \citep[this event has been reported by][]{Nitta2017,He2018}.

At this time, the TDC-scaler subsystem of the HAWC array observed a peculiar GCR flux modulation in the form of a double peak enhancement as  shown in the top panel of Figure \ref{fig:fluxropeSW} 
where HAWC TDC-scaler data: $R_1$,  $M_2$, $M_3$ and $M_4$ are represented. These figures also show the Solar Wind (SW) parameters: V, N (second panel), $B-$field (third panel), the flow and magnetic pressure (fourth panel) as well as the D$_{ST}$ index (bottom panel). We have marked with a dashed vertical red line the arrival time of the shock at 1 AU, whereas the two dotted vertical blue lines mark the starting and ending times of the CME structure at 1 AU as discussed by \citet{Nitta2017} and \citet{He2018}.
The ICME boundaries were selected based on the signatures described on \citet{2006SSRv..123...31Z} and \citet{2010SoPh..264..189R}. There is a  shock wave on 2016 October 12 at 22:01 UT and a sheath that last for 7.68 hr. The MFR boundaries are very well defined with the magnetic field and plasma parameters, starting on 2016 October 13 at 06:00 UT and ending on 2016 October 14 at 16:00 UT with a bulk speed of 390~km~s$^{-1}$. 

Due to the well-defined magnetic structure with clear rotation of B$_Y$ and B$_Z$ (in-situ measurements), the slow CME has been identified as an MC or MFR with a strong southward magnetic field \citep{He2018}. As stated above,  this MFR was the result of the expulsion of a quiet filament with no obvious eruptive signatures in remote sensing observations \citep{Nitta2017}. At 1 AU, the conditions that give rise to the increase in GCRs started with a pre-event HSS (October 1) sweeping the interplanetary medium and imposing very quiet SW conditions ($V \sim 350$ km s$^{-1}$ and $N \sim 7$ part/cm$^3$) in front of the ICME. This smooth SW joined with the low ICME speed ($\sim 450$ km s$^{-1}$) lead to a weak shock and small compression region (sheath, marked with the red and blue vertical lines around October 13 in Figure~\ref{fig:fluxropeSW}). The magnetic pressure (P$_{MAG}\, [nPa] = (10^{-2}/ 8 \pi) \, B [nT]^2$, black) and the SW flow pressure  (computed as P$_{FLOW} \, [nPa] = 2 \times 10^{-6} N [part \,cm^{-3}] \, V [km \, s^{-1}]^2 $, red) are shown in the fourth panel of Figure~\ref{fig:fluxropeSW}, to make it clear that this slow ICME 
is related to a strong magnetic field disturbance. It can be seen that within the MFR (between the two dotted vertical blue lines), P$_{MAG}$ increases while P$_{FLOW}$ decreases, both considerably in comparison to the pressure conditions before the arrival of the MFR (i.e. before the dashed vertical red line). Moreover, the MFR arrival at the Earth produced a relatively intense geomagnetic storm reaching a minimum D$_{ST}$ index of $\sim$ -104 nT \citep{He2018} as shown in the bottom panel of the Figure~\ref{fig:fluxropeSW}.

\subsection{Simulations of the ICME transport and surrounding solar wind conditions} \label{sec:sim}

It is important to note that during 2016 October  the solar activity was in the declining phase of Cycle 24, therefore the SW conditions that we simulate correspond to an interval of low activity in the inner Heliosphere. As the CME propagates 3$^{\circ}$ East of the Sun-Earth line and 11$^{\circ}$ north of the ecliptic plane \citep[for further information see Figure 2 and its description from][]{He2018}, we assume that the CME propagates radially and is directed toward the Earth, and therefore we neglect the deflection and rotation effects \citep{Kay2015}  as well as its interaction with the SW magnetic field \citep{Schwenn2005}. Under these assumptions, hydrodynamic codes give reliable insights into the dynamics of CMEs in the SW \citep{Niembro2019}.

The simulation was done using the 2D hydrodynamic, adiabatic, and adaptive grid code YGUAZ\'U-A, developed by \citet{Raga2000} and used previously to simulate the propagation of ICMEs in the SW \citep[see][for the details of the code and its application to ICMEs]{Niembro2019}. We assume a standard SW \citep[e.g.][]{2012sse..book.....K} with a mean molecular weight\footnote{The mean molecular weight refers to the average mass per particle in units of hydrogen atom mass.} $\mu =$ 0.62 to consider a chemical-composition with solar-mass fractions of 0.7 H, 0.28 He and 0.02 of the rest of the elements; a specific heat at constant volume of $\gamma=5/3$, and an initial temperature of 10$^5$ K \citep{2018ApJS..236...41W}.

The SW conditions at the injection radius R$_{inj}=$ 6 R$_{\odot}$ (4.2 $\times$ 10$^6$ km $\sim$ 0.028 AU) were estimated by splitting the SW measurements at 1 AU into different time periods i = A, B, C, ..., and J, in which the SW showed a speed with a relatively stationary behavior. With this, we estimated the median values of the speed $\overline{V_{1AU_i}}$ and the number density $\overline{N_{1AU_i}}$. Each period is characterized with these median values. Then, we computed the mass-loss rate as $\dot{M_i}=  4 \mu \, m_p \, \pi \,  R_{1AU}^2 \, \overline{N_{1AU_i}} \, \overline{V_{1AU_i}}$ with the proton mass $m_p = 1.67 \times 10^{-27}$ kg. With all these parameters obtained from the measurements at 1 AU, we obtained the speed V$_{inj_i}$ at the injection point R$_{inj}$ (near the Sun) and injection time $\Delta t_i$ by solving the system of equations described in \citet{Canto2005}. All these values are shown in Table \ref{tabla:sw}, with the MFR corresponding to period C.  The computational domain is filled with the conditions described in period A (corresponding to the ambient SW conditions) and the initial injection time is given by the model using the  SW speed observed within period B. 

\begin{table}[htbp] 
\begin{center}
\caption{Median values of the SW speed $\overline{V_{1AU}}$ and number density $\overline{N_{1AU}}$ as extracted from OMNI data set and the computed mass-loss rate $\dot{M}$ used to reconstructed the SW conditions (V$_{inj}$, $\Delta$t) near the Sun at the injection radius R$_{inj}=$ 6 R$_{\odot}$ obtained by solving the system of equations described in \citet{Canto2005}. We have tagged with different colors the periods to be identified and followed in Figures \ref{fig:fluxropeSW} and \ref{fig:sim_evolution}.}
\label{tabla:sw}
\begin{tabular}{|c|c|c|c|c|c||c|c|} 
\hline
 \multicolumn{6}{|c||}{At 1 AU} &
 \multicolumn{2}{|c|}{Near the Sun} \\
 \multicolumn{6}{|c||}{(from OMNI data)} &
 \multicolumn{2}{|c|}{\citep[from][]{Canto2005}} \\
\hline
Period & Initial Date$^{[1]}$ & End Date$^{[1]}$ & $\overline{V_{1AU}}$  & $\overline{N_{1AU}}$ & $\dot{M}$ $\times$ 10$^{-16}$ & $\Delta$t & V$_{inj}$  \\
&  &  & [km s$^{-1}$] & [cm$^{-3}$] & [M$_{\odot}$ y$^{-1}$] & [hrs] & [km s$^{-1}$]  \\
\hline
\colorbox{blue!40}{{\bf A}} & 10/12/2016 00:00 & 10/12/2016 11:55 & 372 & 5.86  & 2.78 & * & 363 \\
\colorbox{green!40}{{\bf B}} & 10/12/2016 20:35 & 10/12/2016 22:15 & 340 & 4.82 & 1.23 & 29.32 & 330 \\
\colorbox{red!40}{{\bf C}} & 10/12/2016 23:18 & 10/13/2016 11:08 & 431 & 17.4 & 5.61 & 11.83 & 423 \\
\colorbox{orange!40}{{\bf D}} & 10/14/2016 08:48 & 10/14/2016 16:08 & 360 & 7.49 & 2.02 & 16.5 & 351 \\
\colorbox{yellow!40}{{\bf E}} & 10/14/2016 16:53 & 10/14/2016 18:33 & 405 & 5.53 & 1.68 & 1.66 & 397 \\
\colorbox{blue!40}{{\bf F}} & 10/14/2016 21:33 & 10/14/2016 23:23 & 358 & 10.96 & 2.94 & 16.79 & 349 \\
\colorbox{green!40}{{\bf G}} & 10/15/2016 01:03 & 10/16/2016 06:30 & 502 & 4.51 & 1.69 & 66.71 & 495 \\
\colorbox{red!40}{{\bf H}} & 10/16/2016 08:43 & 10/16/2016 17:03 & 665 & 3.87 & 1.93 & 44.05 & 660 \\
\colorbox{orange!40}{{\bf I}} & 10/16/2016 18:58 & 10/18/2016 04:18$^{**}$ & 705 & 2.32 & 2.25 & 56.38 & 700 \\
\hline
\end{tabular} \\
\end{center}
[1] [mm/dd/yyyy hh:mm UT] \\
* Period A corresponds to the SW conditions needed to fill in the computational domain in which the different periods will be injected. \\
** The end of the period is not shown in Figures \ref{fig:fluxropeSW} and \ref{fig:sim_evolution}.
\end{table}

The synthetic speed and number density profiles resulting from the simulation are shown as orange solid lines in the second panel of Figure~\ref{fig:fluxropeSW} to be compared with  the in-situ measurements at 1 AU (shown in the figure in black and green, respectively). It can be noticed that our approach accurately resembles the observed speed profile, while the  number density profile is consistent with the observations, except for those time ranges in which the 3D-configuration of the magnetic field certainly plays a major role in the dynamics \citep[e.g.][]{2002AnGeo..20..879C}, such as in the compression region formed due to the interaction of the MFR with the ambient SW in front of it and delimited by the dashed red vertical line (shock) and the first dotted blue vertical line (initial time of the MFR). The other two enhancements found at later times in the number density profile (2016 October 14 $\sim$17:00 UT and 2016 October 16 $\sim$12:00 UT) are related to compression regions formed by the interaction between two flows at different speed  (between periods F and G and, G and H, respectively). It is worth noting that from the hydrodynamics point of view,
the plasma is compressed when two flows at different speeds interact, while a low-density region will be formed when the flows separate from each other.

\begin{figure}[htpb]
\includegraphics[width=\textwidth]{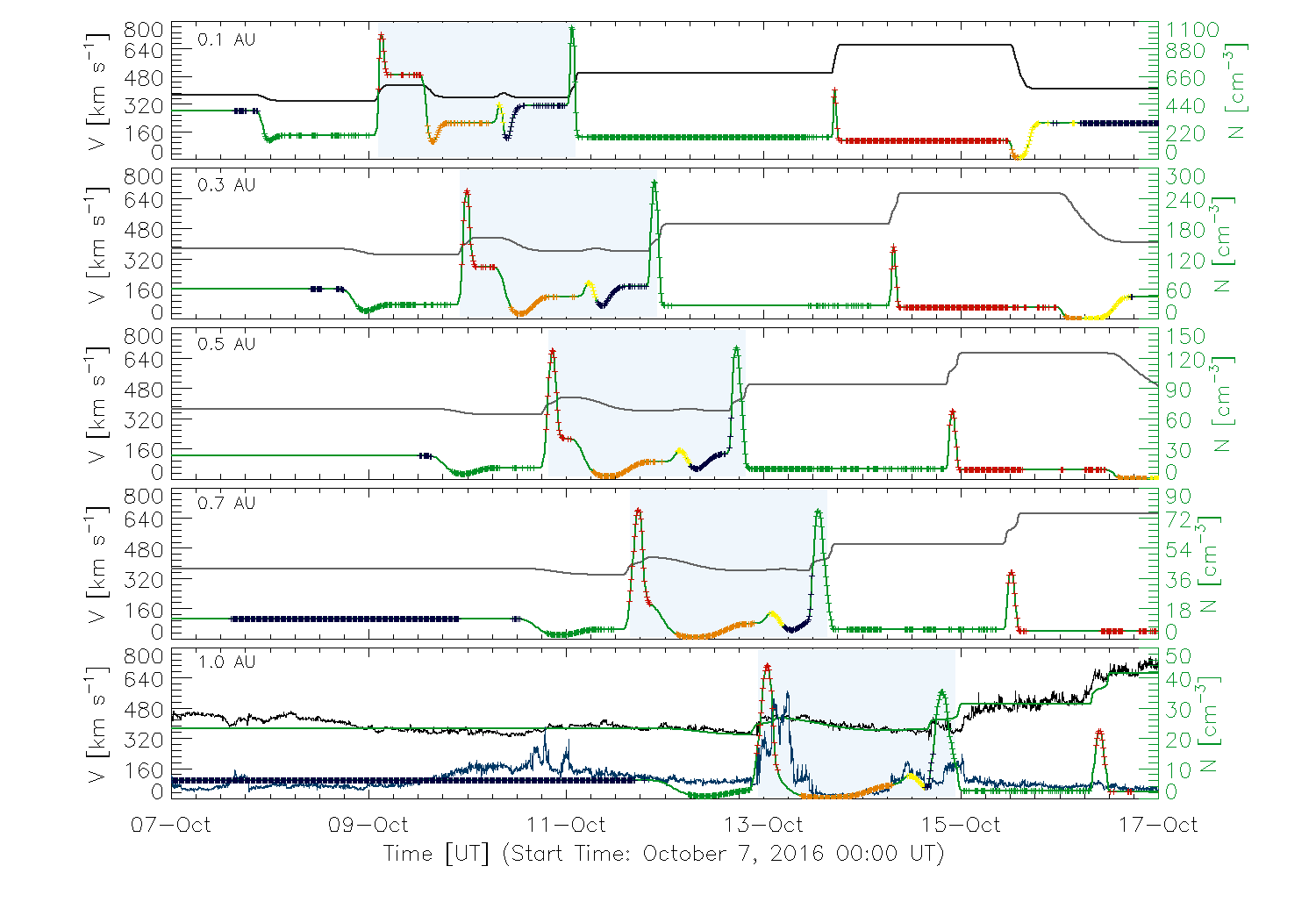}
\caption{The synthetic speed (black solid line) and number density (green solid line) profiles obtained by simulating the variable SW conditions with YGUAZ\'U-A at five different heliospheric distances (from top to bottom): 0.1 AU, 0.3 AU, 0.5 AU, 0.7 AU and 1.0 AU. The colored symbols tag the different plasma parcels as listed on Table~\ref{tabla:sw}. The duration of the CME (48 hrs delimited by the two maxima density enhancements shown in red and green respectively) is marked within a blue shaded area in each panel to show that it does not change during the evolution, the density cavities prevented the interaction of the CME with the surrounding SW conserving its structure. In the bottom panel  we show the comparison between the observed and synthetic SW speed (black and green lines respectively) and number density (blue and green lines respectively) at 1 AU.}
\label{fig:sim_evolution}
\end{figure}

An important feature of this particular hydrodynamic code is that it enables us to tag and follow each different plasma parcel. This is shown with the colored symbols used over the synthetic number density profile which are indicated in Figure~\ref{fig:fluxropeSW} following the slightly lighter color code in Table~\ref{tabla:sw}. We would like to focus on the evolution of two of the simulated SW parcels colored in green (corresponding to period B) and in orange (period D). It can be noticed that the SW delimited in these two periods presents low density cavities which can be easily tracked in Figure~\ref{fig:sim_evolution} where  we show the speed (black solid line) and number density (green solid line with colored symbols) profiles at five different heliospheric distances: 0.1 AU, 0.3 AU, 0.5 AU, 0.7 AU and, 1.0 AU, from top to bottom, respectively.

These low density cavities prevented the interaction of the disturbance with the surrounding ambient SW conserving the well-defined and strong magnetic structure of the MFR from its source up to 1 AU. Moreover, the duration of the CME (48 hrs, time delimited by dashed blue lines in Figure~\ref{fig:fluxropeSW} and blue shaded area in Figure~\ref{fig:sim_evolution}) does not change at different heliospheric distances, thus  allowing  the guiding of GCRs observed by HAWC.

\subsection{Effects of the Magnetic Flux Rope on the Earth's Magnetospheric Field} \label{sec:magnetosphere}

Besides the spacecraft located around the L1 point, there are several spacecraft orbiting the Earth at different locations which are able to detect changes on the Earth's magnetic field caused by the ICME (shock, sheath, and flux rope) in the Earth vicinity. In this work, we use $B-$field data from THEMIS-E and THEMIS-C \citep{Angelopoulos2008}, GOES-15 (\url{https://www.goes.noaa.gov/goesfull.html}) and MMS-1 \citep{Burch2016} to characterize the disturbances caused by the 2016 October MFR on the Earth's magnetosphere and its possible effects on the GCR flux.

\begin{figure}[ht!]
\hspace*{.5cm}
\includegraphics[width=0.9\columnwidth]{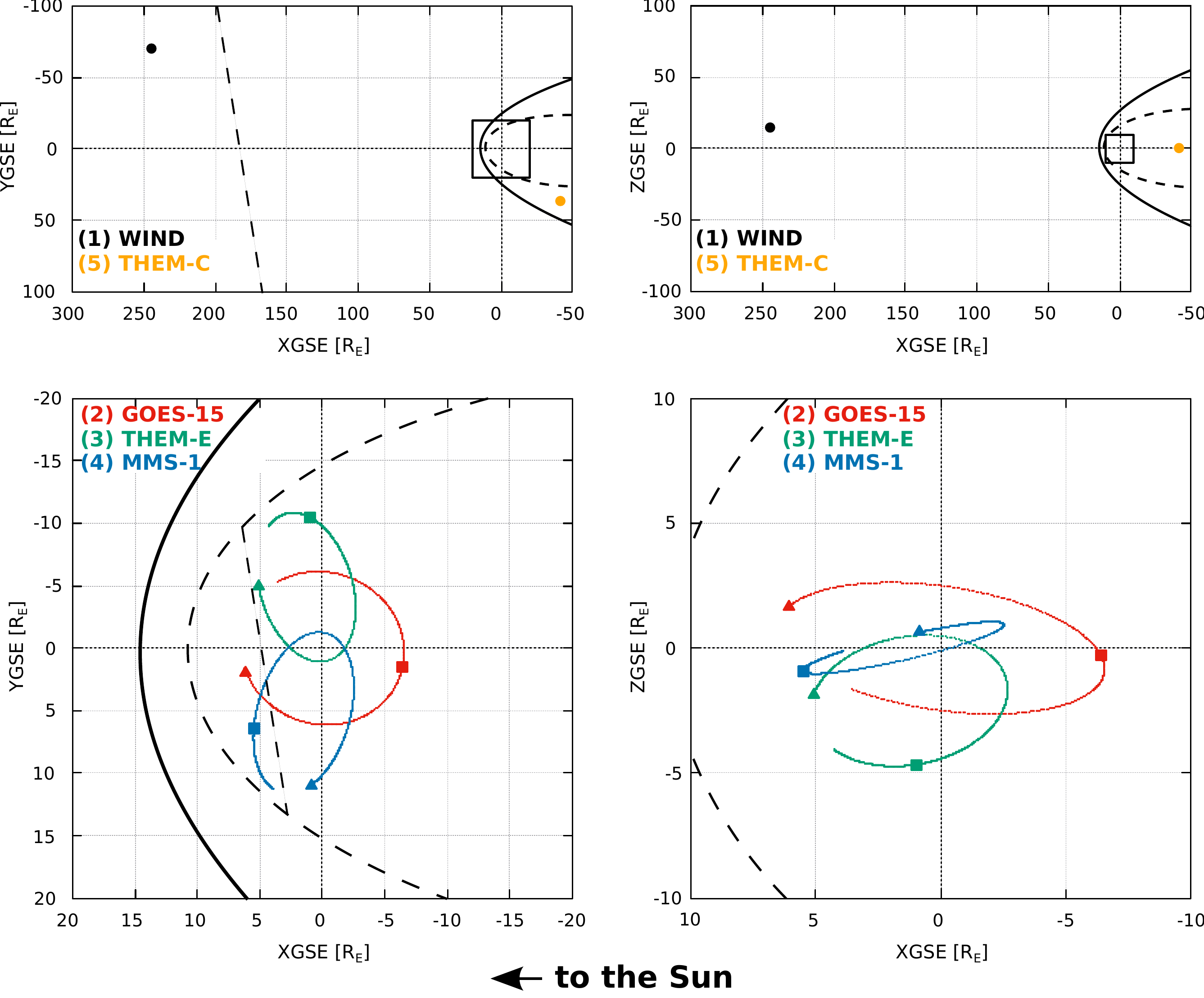} 
\centering
\caption{Spacecraft orbits/positions showing the shock/$B$-discontinuity time (triangle) and $B_Z$ decrements (square) projected on XY (left) and XZ (right) GSE-planes (where $R_E =$ 6357 km is the Earth's radius at the equator). The line-square on the upper panels represent the plotted area in bottom panels. The number between brackets next to the spacecraft name indicates the timing of the shock/$B$-discontinuity observation. A model of bow-shock (continuous line) and magnetopause (dashed line) is shown. The 9$^\circ$ inclined dashed line in XY panels corresponds to the MFR major axis orientation.}
\label{fig:preisserorbits}
\end{figure}

Figure \ref{fig:preisserorbits} shows the orbits of these spacecraft in the $XY$ (left) and $XZ$ (right) GSE planes  (the geocentric GSE is a right-handed coordinate system with $X$ axis pointing toward  the Sun and the $Z$ axis is perpendicular to the ecliptic plane and positive pointing North). The triangles on the panels in Figure \ref{fig:preisserorbits} mark the spacecraft locations at the time when the magnetic field discontinuity associated with the arrival of the ICME driven shock was observed (column 3 in Table \ref{tabla:spcrft} and vertical gray lines in the top panels of Figures~\ref{fig:preisserwind}, \ref{fig:preissergo15}, \ref{fig:preisserthe_e}, \ref{fig:preissermms1} and \ref{fig:preisserthe_c} in Appendix~\ref{app:magnetosphere}). The squares in Figure \ref{fig:preisserorbits} mark the location of the sudden decrease in~$B_Z$ associated with the arrival of the leading edge of the MFR at each spacecraft (column~6 in Table~\ref{tabla:spcrft}). The bow shock boundary drawn (continuous line) was calculated from a modified version of \citet{Fairfield1971} while the magnetopause location (dashed line) is based on the \citet{Roelof1993} model, both boundaries are shown in Figure \ref{fig:preisserorbits} only as reference and are based on the SW average value of $P_{FLOW}=$ 1.65 nPa measured at 1 AU. It is important to note that during the MFR passage these boundaries maybe largely modified. According to this and in agreement with the observations, GOES-15, THEMIS-E and MMS-1 were located inside the magnetosphere during the time interval studied, whereas THEMIS-C was in the magnetosheath region.

\begin{figure}[ht!]
\hspace*{-0.5cm}
\includegraphics[width=0.8\columnwidth]{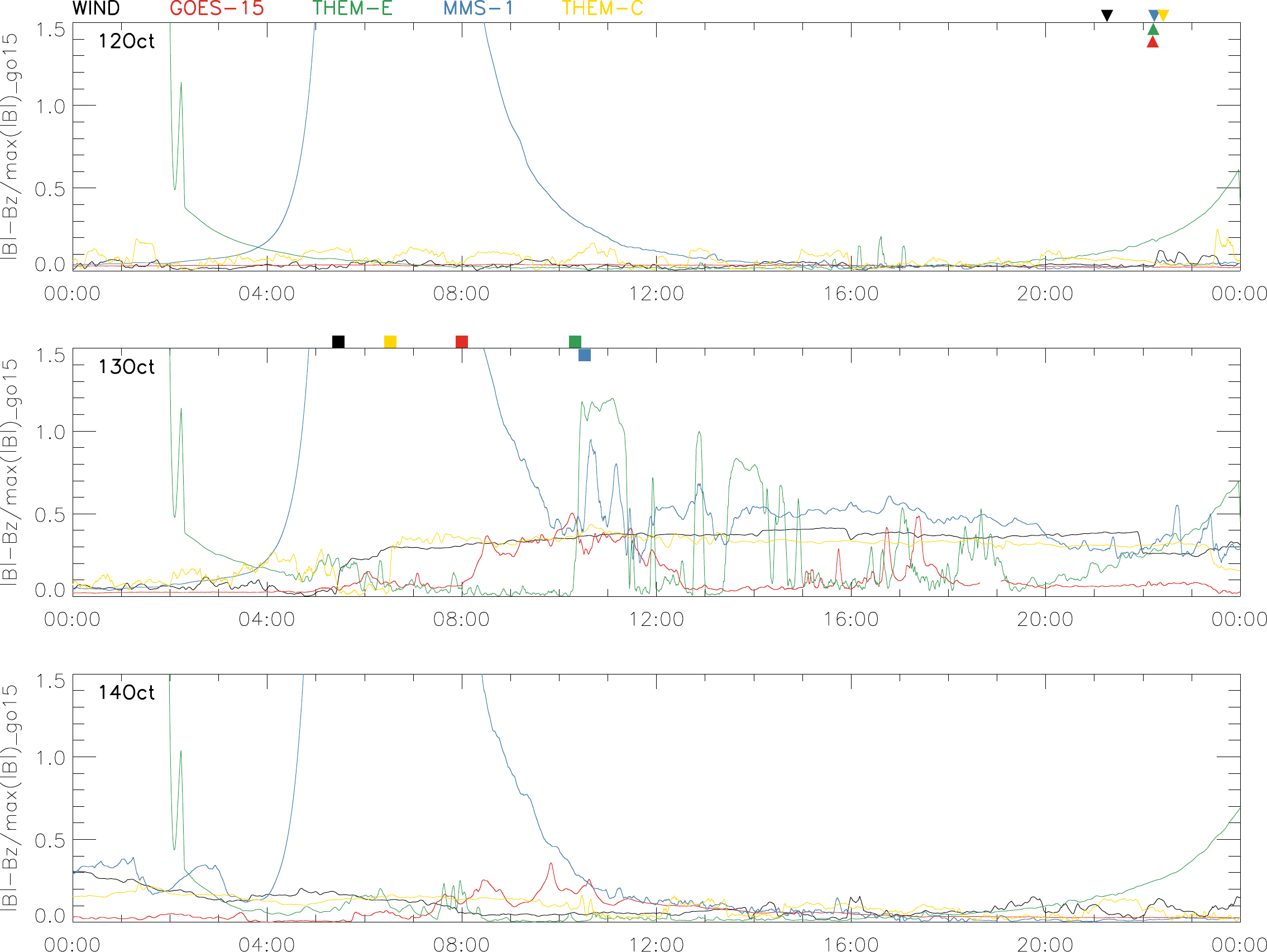} 
\centering
\caption{Plot of $(|B|-B_{Z})/max(|B|)_{GOES-15}$ for all the spacecraft (in different colors) observed on October 12 (top), 13 (middle) and 14 (bottom), 2016. The colored symbols in panels correspond to the timing of $B-$discontinuity (triangles) and $B_Z-$decrement (squares) according to Table \ref{tabla:spcrft} and Figure~\ref{fig:preisserorbits}.
\label{fig:preissermulti}}
\end{figure}

In general, inside the magnetosphere the magnetic field is dominated by the North-South geomagnetic dipole, i.e., $B_Z$ component. Therefore, we follow and compare the values of this $B-$field at different locations around the Earth during the MFR passage. In Figure~\ref{fig:preissermulti}, we show the difference $|B|-B_Z$ normalized to the maximum value of $|B|$ observed by GOES-15 which
of the spacecraft considered in this work, was the one with the shortest and constant distance to the Earth at the time of the sudden decrease in
$B_Z$ sudden decrease. This difference was computed for all the spacecraft (plotted with different colors) during October 12, 13 and 14, 2016 (top, middle, and bottom panels, of Figure~\ref{fig:preissermulti} respectively). The data are in GSE-coordinates except for GOES-15 which is presented in EPN-coordinates  E (Earthward), P (Parallel) and N (Normal) with respect to the spacecraft orbit plane.  (see Appendix~\ref{app:magnetosphere}). The individual plots of $B-$field and its components measured in each spacecraft during the event are shown in Figures~\ref{fig:preisserwind} to \ref{fig:preisserthe_c} in Appendix~\ref{app:magnetosphere}.

\begin{figure}[ht!]
\includegraphics[width=0.8\columnwidth]{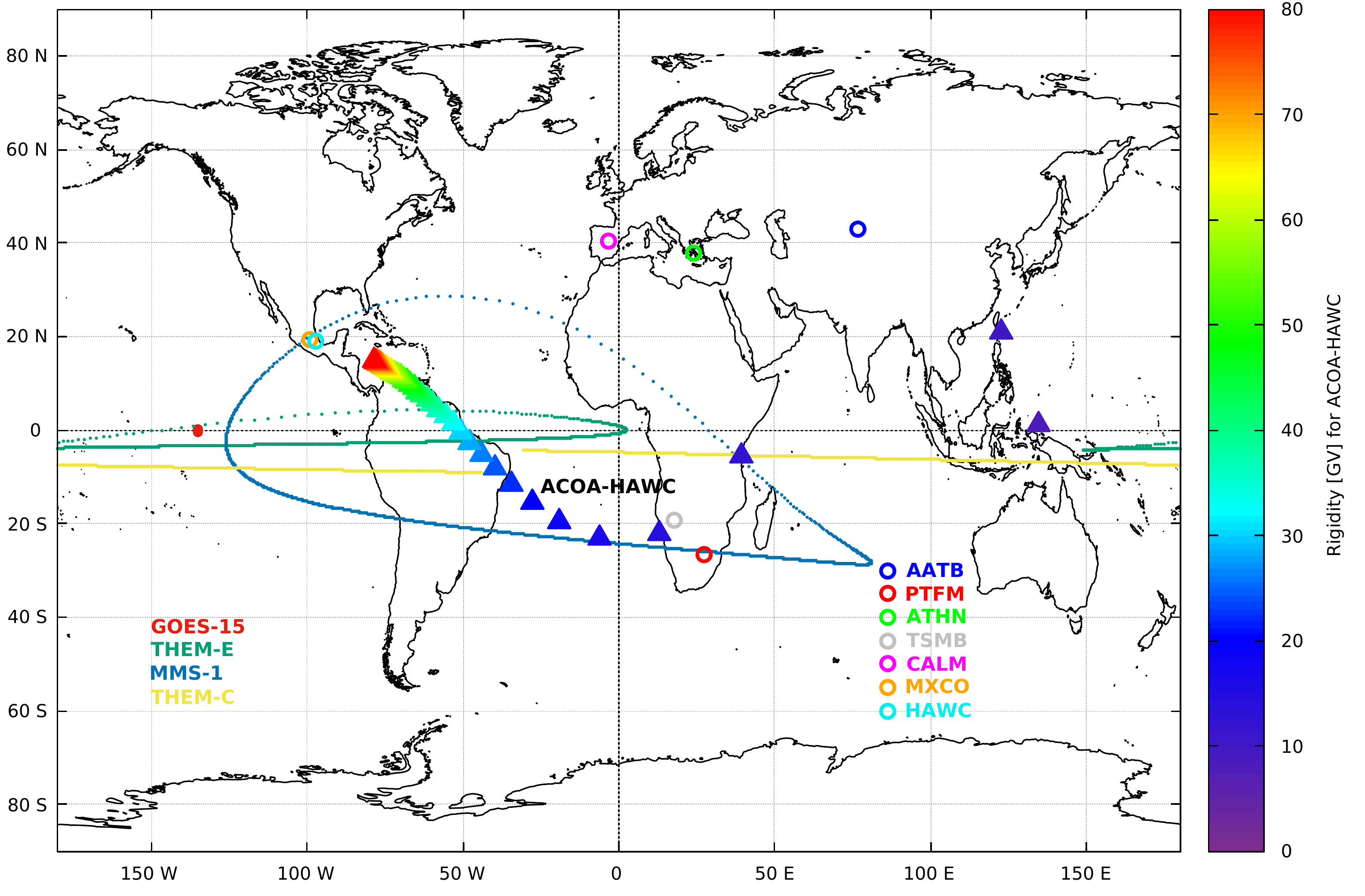} 
\centering
\caption{GEO projection of spacecraft paths (continuous and dotted lines) corresponding to October 13, 2016 and location of Neutron Monitors (open-colored circles). The rainbow-triangle-path crossing the equator line indicates the rigidity values (shown in color palette) for the Asymptotic Cone of Acceptance for HAWC observatory  (marked as ACOA-HAWC).  
\label{fig:preissergeo}}
\end{figure}

In Table \ref{tabla:spcrft}, we summarize the times at which the spacecraft observed the shock-associated discontinuity of $B-$field and their corresponding upstream/downstream $B-$values (columns 3, 4 and 5, respectively), the times of the sudden decrements of $B_Z$ associated to the MFR and its minimum value (columns 6 and 7) as well as the magnetospheric location (column 8) of each analyzed spacecraft (column 2). 
The Geo projection of the spacecraft during the MFR passage are plotted in Figure \ref{fig:preissergeo}.

As the MFR enters to the magnetospheric region, following the ICME sheath region, the decrease in $P_{FLOW}$ (see fourth panel in Figure~\ref{fig:fluxropeSW}) associated to the internal region of the MFR produced an expansion of the Earth's magnetic field. This effect can be seen by comparing the October 13 time intervals (04:00-08:00 UT) in Figure~\ref{fig:preissergo15}, (04:00-09:00 UT) in Figure \ref{fig:preisserthe_e} and (08:00-10:00 UT) in Figure \ref{fig:preissermms1} in Appendix \ref{app:magnetosphere} with the same time period but during the previous day. This effect is clearer for GOES-15 in Figure~\ref{fig:preissergo15} due to its geostationary orbit.

This magnetospheric weakening, along with the enhancement of the southward magnetic field component inside the MFR, caused sudden reversals ($B_{Z}<$~0) on the Earth's magnetic field observed as step-like signatures by THEMIS-C, THEMIS-E and MMS-1 as shown in the middle panel of Figure \ref{fig:preissermulti} and as changes in the polarity of $B_Z$ in Figures~\ref{fig:preisserthe_e}, \ref{fig:preissermms1} and \ref{fig:preisserthe_c} in Appendix \ref{app:magnetosphere}. 

Because of the unique position of GOES-15, in particular its proximity to the ecliptic plane and its low altitude, the  intense local magnetic field makes it difficult to clearly detect the perturbation caused by the MFR, as shown in Figure~\ref{fig:preissermulti} and in Figure~\ref{fig:preissergo15} in the Appendix \ref{app:magnetosphere}. However, significant disturbances were observed during most of October 13 and continue but with lower amplitude the day after, when the GCR double peak was observed by HAWC,  as can be corroborated observing the middle and bottom panels on Figure \ref{fig:preissermulti}, showing in this way, that the GCR enhancement observed by HAWC was not directly caused by these geomagnetic disturbances.

\begin{table}[htbp]
\definecolor{go15}{HTML}{E51E10}
\definecolor{thee}{HTML}{009E73}
\definecolor{mms1}{HTML}{0072B2}
\definecolor{thec}{HTML}{FFA500}
\begin{center}
\caption{Parameters related to the spacecraft observations.} 
\label{tabla:spcrft}
\begin{tabular}{|c|c|c|c|c|c|c|c|c|c|c|}
\hline
$s/d$ & Spacecraft & $t_{s/d}$ [UT]      & $B_{up}$ & $B_{down}$ &  $t_{B_Z}$       & min($B_Z$) & Spacecraft \\
      &           & DD/MMM hh:mm:ss & [nT]  & [nT]  &  DD/MMM hh:mm:ss & [nT]       & Region*   \\
\hline
s & \textcolor{black}{\textbf{WIND}} & 12/OCT 21:15:37 & 3.4 & 8.2 & 13/OCT 05:27:30 & -20.0 &SW\\  
d & \textcolor{go15}{\textbf{GOES-15}} & 12/OCT 22:11:51 & 121 & 150 & 13/OCT 08:00:00 & 8.0 &M-SPH\\  
d & \textcolor{thee}{\textbf{THEM-E}} & 12/OCT 22:13:33 & 92.5 & 123 & 13/OCT 10:19:21 & -60.0 &M-SPH\\
d & \textcolor{mms1}{\textbf{MMS-1}} & 12/OCT 22:13:52 & 30 & 47 & 13/OCT 10:31:10 & -50.0 &M-SPH\\ 
s & \textcolor{thec}{\textbf{THEM-C}} & 12/OCT 22:24:49 & 6 & 10.5 & 13/OCT 06:31:37 & -22.5 &M-SH\\
\hline
\end{tabular} \\
* s/d: shock or discontinuity, SW: solar wind, up: upstream, down: downstream, M-SPH: magneto-sphere, M-SH: magneto-sheath
\end{center}
\end{table}

\subsection{Ground-level observations by HAWC and Neutron Monitors}

\begin{figure}[ht!]
\centering
\includegraphics[width=0.8\columnwidth]{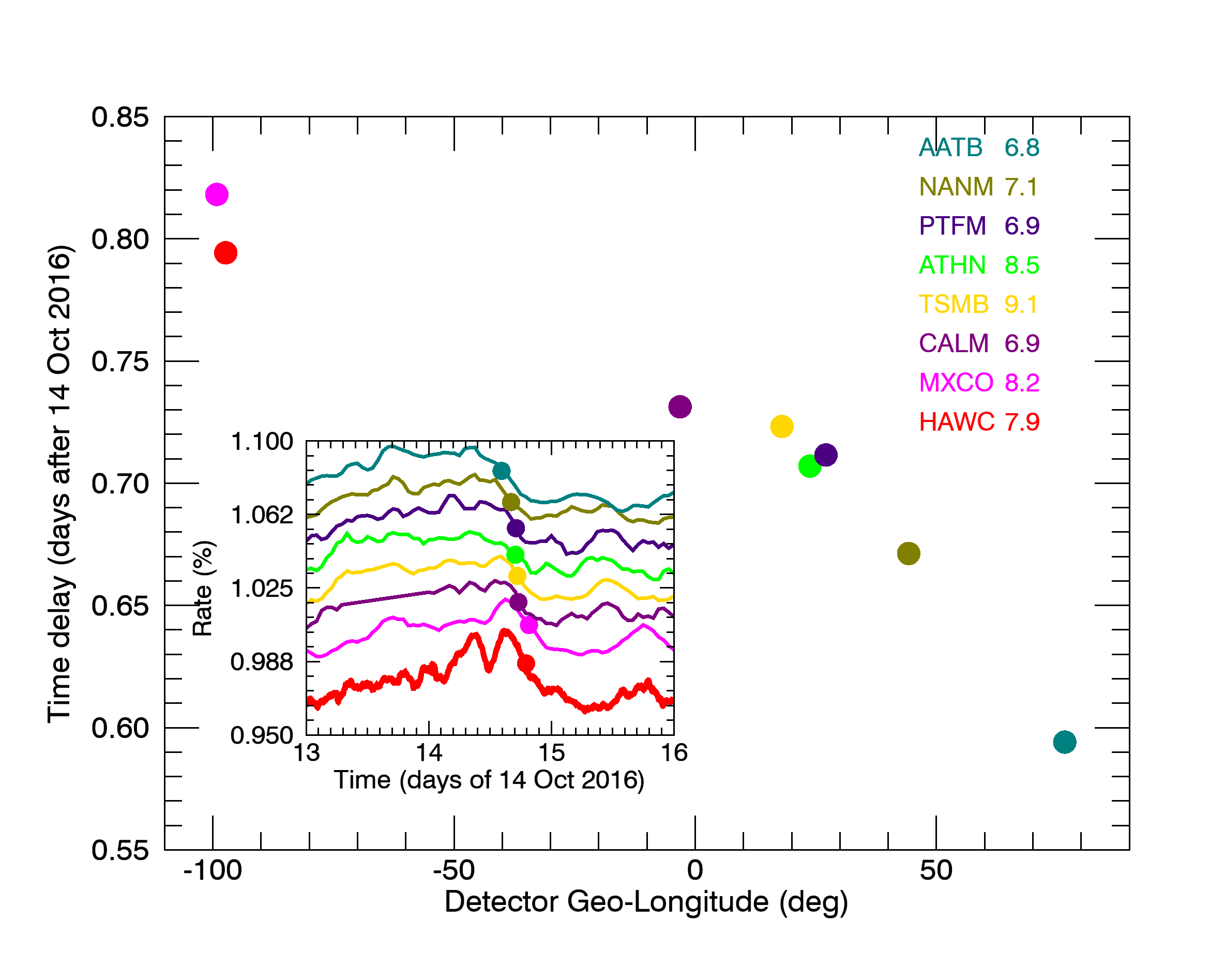}
\caption{The MFR passage observed by different NM around the world, whose cutoff rigidities  [GV] are similar to those  of HAWC (the acronym and cutoff rigidity of each monitor is in the upper right corner). The  outer frame shows the delay of the perturbation decrease as a function of the Geo-longitude of each NM. The inner frame shows the time profiles of the NM as well as the M3 HAWC rate  divided by 30.} 
\label{fig:nm}
\end{figure}

Although the GCR anisotropy induced by the MFR must be a global event, the response of Neutron Monitors (NMs) with cutoff rigidities similar to that of HAWC (6 - 9 GV), during the passage of the MFR was somewhat poor due to the fact that the sensitivity of NMs is lower than that of HAWC in this energy range.  These responses are shown in the inset  of Figure~\ref{fig:nm} where the rates of the selected NMs during a three-day period, starting on 2016 October 13 are shown. For the sake of clarity, we have added/subtracted an offset to each time profile and the M3 rate of HAWC (red line) is divided by 30 to fit the scale.  

Even though the behavior of the time profile is different for each monitor, all of them show an increasing trend after the day 13 and a more evident decrease after the day 14.5. It is interesting to note that the Mexico City NM (magenta line), has a time profile similar to HAWC during and after the second  peak  which was caused by low-energy particles (see Section \ref{sec:model}). The first peak as mentioned before was originated by high-rigidity ($>$ 15 GV) protons for which the NM response functions are not well determined \citep{Lockwood1971}. 
 
 Due to the fact that neither the initial nor the peak times can be precisely determined, we use the well observed decrease after the second peak, to determine the final time of the perturbation on each NM, computed  as the mid-time between the maximum and the first change of curvature of the decreasing trend. These times are marked with circles in Figure~\ref{fig:nm} on the time profiles (inset). The main plot  shows the time delay (the fraction of a day after 2016 October 14) of the decrease observed by each NM as a function of its Geo-longitude, and shows a clear relationship between the position of NM and the final time of the observed interplanetary disturbance. This nonlinear relation is caused by multiple factors like Earth's rotation, solar wind velocity at which MFR is sweeping, the energy of observation, and the detector response function. More detailed study of this effect is required but is out of scope of this paper. 
 
\section{Flux-rope Model and the GCR anisotropy}  
\label{sec:fr_gcr}
\subsection{Fitted MFR Model} 
\label{sec:frmodel}

Based on visual inspection, the magnetic configuration of the MFR displays a symmetric magnetic field profile. The structure can be described with a very well-organized single MFR with a south–north (SN) $B_Z$ polarity and positive $B_Y$. Thereby, the configuration is defined as a south-east-north configuration and left-handed (SEN-LH) \citep[see][and references there in for  details]{2019SoPh..294...89N}.
The MFR reconstruction is based on the circular-cylindrical model described by \citet{2016ApJ...823...27N}. The model assumes an axially symmetric magnetic field cylinder with twisted magnetic field lines of circular cross section. The magnetic field components in the circular-cylindrical coordinate system are described by:

\begin{eqnarray}
    B_r &=& 0, \\
    B_y &=& B_y^0 \left[1- \left(\frac{r}{R} \right)^2 \right], \\
    B_\phi &=& -H \frac{B_y^0}{|C_{10}|}\frac{r}{R},
\end{eqnarray}

where the model estimates $B_y^0$ , the magnetic field at the center of the MFR, and $C_{10}$, a measure of the force-freeness along with handedness $H$ (+1 or $-$1)  that indicates whether the MFR is right- or left-handed. The radius of the MFR cross section, $R$, is a derived parameter \citep[see Equation 5  in][]{2002JGRA..107.1002H} and $r$ is the radial distance from the axis that describes the spacecraft trajectory. The reconstruction technique is based on a multiple regression technique (Levenberg-Marquardt algorithm), which infers the spacecraft trajectory and estimates the MFR axis orientation (azimuth and tilt), and the impact parameter (y$_0$).

The reconstruction of the event corroborates the visual inspection,  i.e., the structure corresponds to a symmetric MFR with an axis orientation of $\phi  = 99^\circ$  and $\theta = -21^\circ$ of longitude and latitude. The closest distance of the spacecraft to the axis was $y_0 = -0.0054$ AU and the radius $R= 0.146$ AU, therefore crossing very close to the MFR center ($y_0/R \approx 0.04$). The $H$ is negative, left-handed. The magnetic field strength at the MFR axis is $B_y^0 = 20.7$ nT and the force-freeness parameter C$_{10}=$ 1.1 \citep[see][for more details]{2016ApJ...823...27N}.

\begin{figure*}[htbp]
\includegraphics[width=0.9\columnwidth]{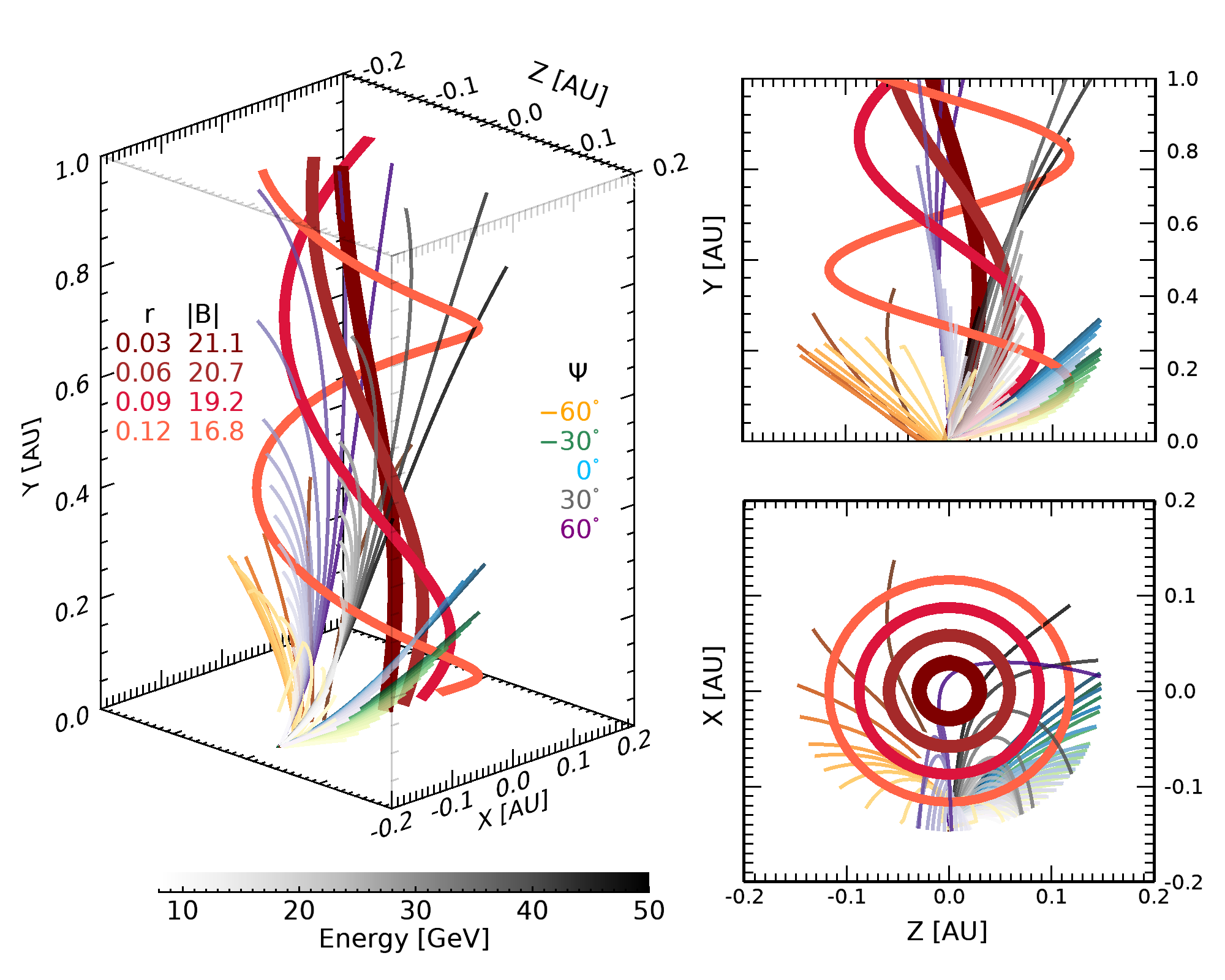}
\caption{Three views of the fitted MFR. In reddish colors we show four magnetic lines of the fitted MFR at the radius [AU] and with the magnetic field strengths [nT] annotated in the left side of the left panel. We have simulated the trajectories of GCRs inside this fitted MFR and show some examples at different incidence angles ($\Psi$, as described in the left panel). The shades of colors represent the GCR energy as shown by the color bar in the bottom left side. The right panels show the MFR projections in the ZY (upper) and XZ (bottom) planes (see Section~\ref{sec:model} for details).}
\label{fig:trayectories}
\end{figure*}

The twisted reddish curves in Figure~\ref{fig:trayectories} represent the fitted MFR at four  radial distances from its axis, as quoted in the left side panel. As described in Section~\ref{sec:model}, we used these fitted parameters to model the trajectories (inside the MFR) of the GCRs with different energies and incident angles. The colored curves starting at $x=-0.14$ AU, $y=0$ AU and $z=0$ AU, represent these GCR trajectories.

\subsection{GCR guiding inside the MFR}
\label{sec:model}

\begin{figure}[ht!]
\includegraphics[width=0.8\textwidth]{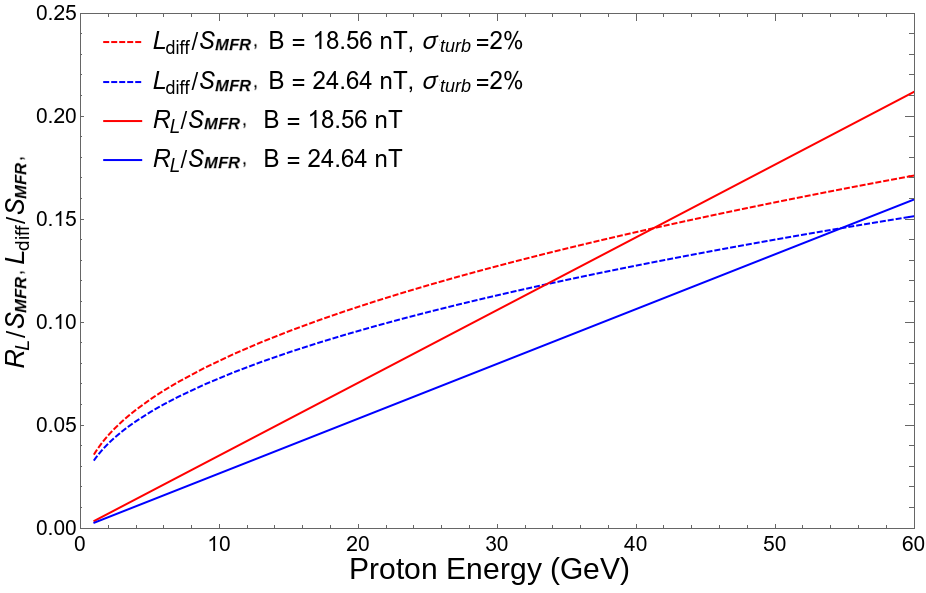}\centering
\caption{The solid lines show the ratio between the Larmor radius  ($R_L$) and the MFR size, whereas the dashed lines show the ratio between diffusion length and the MFR size for the maximum (blue) and mean (red) magnetic field magnitudes with magnetic turbulence level $\sigma_{turb} = 2\%$.}
\label{val_pt}
\end{figure}

As shown in the previous section, 
the 2016 October  ICME had a well-defined magnetic rope topology, ideal for a charged particle to be guided along the MFR axis (governed by the Lorentz force). The $B-$field inside the MFR reached a maximum of $\sim 24.64$ nT, with a mean value ($<B>$) of $\sim 18.56$ nT. A GCR entering the MFR experiences a Lorentz force which is maximum while it travels perpendicular to the axis and minimum when its direction is parallel to the axis. The ratio of Larmor radius ($R_L$) over the MFR size ($S_{MFR}$) is shown in Figure~\ref{val_pt}, the low value of this ratio indicates the field capacity to redirect the particles and therefore increase the likelihood of an alignment of the GCRs along the $B-$field (note that this is the field strength along the axis of MFR). 

Furthermore, it should be noted that the $B-$field associated with this MFR had a low turbulence level, which was estimated as $\sigma_{turb} \leq2\%$, using a running window method \citep[as explained by][]{Arun2015}. The turbulence level enhances the diffusion of particles, but in this event the turbulence level was low. To quantify this effect, we estimated the ratio of diffusion length ($L_{diff}$) of the particles to $S_{MFR}$ using the perpendicular diffusion coefficient described in \citet{Snodin2016} with a turbulence level of $\sigma_{turb} = 2\%$. The lower values of this ratio of $L_{diff} / S_{MFR}$ as seen in Figure~\ref{val_pt} supports our approximation and illustrates the viability of the model.

To model the trajectory of GCRs inside the MFR we define a coordinate system with the origin at the center of the MFR and the Y-axis aligned with the MFR-axis. This can be converted from the GSE-coordinate system by the rotation operators $R_Z$ and $R_Y$ by the angles  $\theta$ and $\phi$ (see Section \ref{subsec:ICME}), respectively. We assumed the MFR has a cylindrical cross-section, where the boundary of the MFR lies in the XZ-plane with $\sqrt{x^2+z^2} \leq 0.5  S_{MFR}$. The field inside the MFR was modeled using the observations at 1 AU and then converted to the coordinate system based on the MFR-geometry.

The GCRs entering the MFR are redirected by the Lorentz force ($\frac{q}{m_p} \vec{V} \times \vec{B}$). The net deflection is then transformed (relativistically) into the particle frame. When the particle is inside the MFR, its changes in position and velocity are estimated every 100 m of travel. Figure \ref{fig:trayectories} shows few examples of the simulated particle trajectories inside the fitted MFR.

To estimate the relevant energies of particles that can be guided along the axis of the MFR we performed a simulation, for which we injected protons with energies ranging from 8 to 100 GeV entering to the MFR at an initial position $x= -0.5 \times S_{MFR}$, $y=0$ and $z=0$. These particles enter the MFR with an incident angle in the XY-plane $\Psi$, ranging from  -$70^{\circ}$ to $70^{\circ}$ with an increment of $1^{\circ}$. We tracked the position and direction of each particle inside the MFR by computing the point where it becomes aligned with the axial direction of the MFR (inside the $85^{\circ} < \Psi < 95^{\circ}$ cone) and the distance it travels parallel to the axial direction. We found that that a rather narrow range of rigidity behaves in this manner, specifically, the particles of energies higher than $\sim$ 60 GeV are less likely to be aligned along the axis (see Appendix~\ref{app:axial} for details). Therefore, the enhancement observed by HAWC is due to particles of energies $\lesssim$ 60 GeV in this model.

As an example, in Figure~\ref{fig:trayectories} we show the trajectories inside the MFR of protons with energies between 8 and 50 GeV (marked with color shades) entering the MFR with five different incident angles within the range $-60^{\circ} < \Psi < 60^{\circ}$ and an increment of  $ 30^{\circ}$ (marked by the different colors). The entrance position of this example was $x= -0.5 \times S_{MFR}$, $y=0$ and $z=0$. The incident angle $\Psi=0$ means that the particle is entering the MFR perpendicular to the Y-axis and the initial velocity of the particle was resolved into perpendicular and parallel components with respect to the MFR axis, using this incident angle as $V_{\perp} = V ~ \cos \Psi$ and $V_{\|} = V ~ \sin \Psi$. As expected, the low-energy particles are more affected by the magnetic topology of the MFR. Even though the median energy particles with larger $\Psi$ are more likely to get aligned parallel to the MFR-axial direction.

\begin{figure*}[ht!]
\centering
\includegraphics[width=0.9\columnwidth]{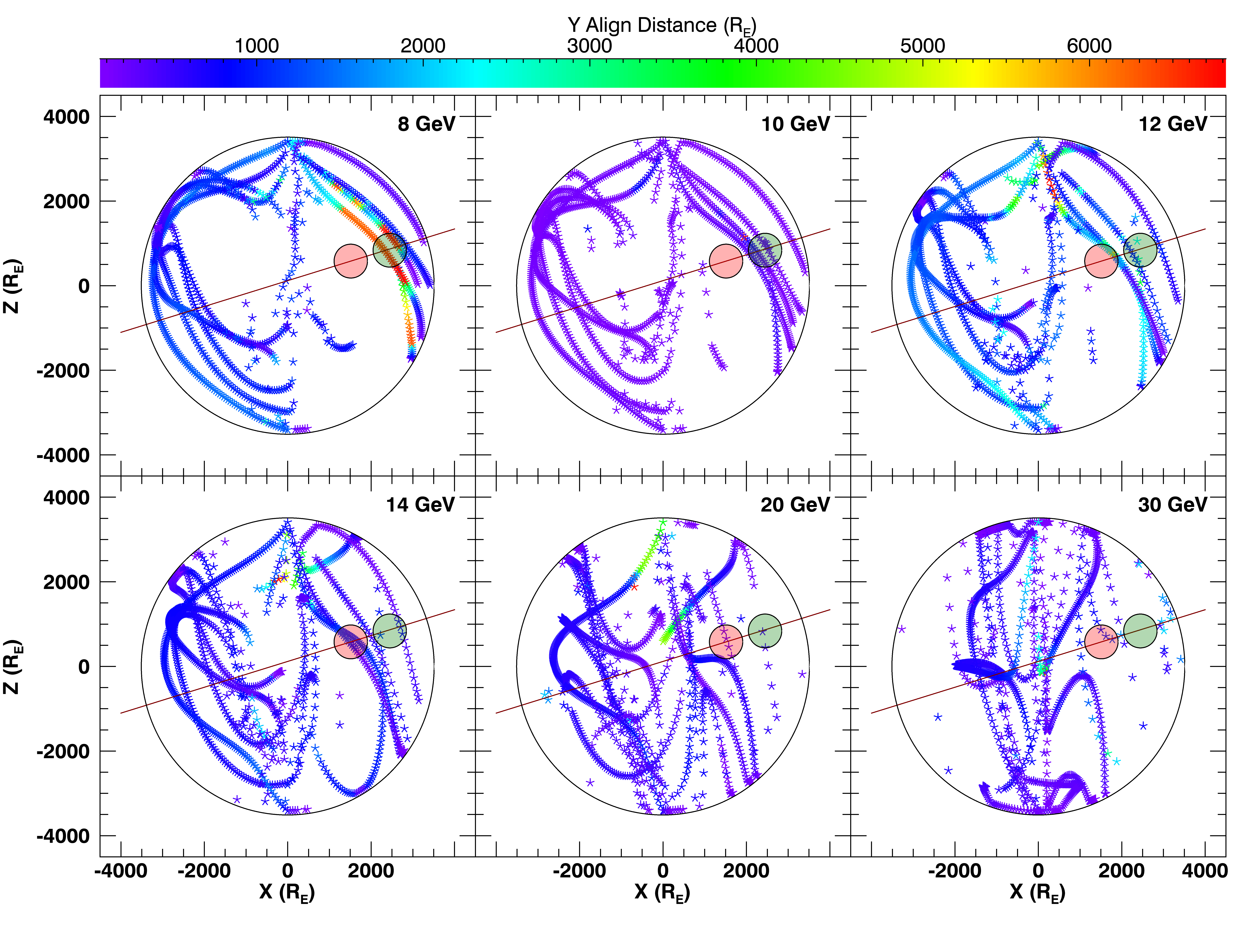}
\caption{Projections on the plane XZ of the simulated GCR trajectories inside the MFR, each panel corresponds to different energy particles (as marked in the top right corner of the panel). 
Here each point shows the position in XY cross-section where it started getting aligned in the axial direction, also the color code shows  the distance traveled by the particles  parallel to the MFR axis. The brown lines show the path of the Earth through the MFR  transiting from left to right and the red and green circles correspond to the approximate duration of the first and second peak of the HAWC rate excess, respectively.} 
\label{crosec}
\end{figure*}

For completeness and to cover the maximum possible entries of particles into the MFR, we introduced another angle of incidence $\Xi$ surrounding the MFR in the XZ-plane, varying from $1^{\circ}$ to $360^{\circ}$  in increments of  $1^{\circ}$ and  the initial position of entry chosen as $ x = -(0.5\times S_{MFR})~ \cos \Xi$, $y=0$ and $ z= -(0.5\times S_{MFR})~ \sin \Xi$. If the particles get aligned and travel along the axial direction for more than 60 $R_E$ (with $R_E =$ 6357 km, the Earth's radius at the equator), the initial position of alignment and the distance traveled  parallel to the MFR axis  are estimated. The projections of the trajectory of these particles (in the XZ-plane, i. e. in the MFR cross-section), with initial energies of 8, 10, 12, 14, 20, and 30 GeV are shown in Figure~\ref{crosec}. The distance traveled parallel to the MFR axis is indicated by the color scale. In this figure, the trajectory of the Earth through the MFR is marked by the brown lines whereas the red and green circles mark the times when the double peak structure was observed by HAWC. Therefore, the number of points inside these circles shows the trajectory of the particles that are likely  to be heading toward the Earth at these times. As seen in this figure, our model suggests that the first enhancement detected by HAWC was mainly caused by protons in the energy range of 12-30 GeV, whereas the second enhancement was caused by low-energy protons in the 8-12 GeV range.

To reach the lower atmosphere and be detected by HAWC, the GCR anisotropy caused by the MFR (parallel to its axis) must be matched by the HAWC asymptotic directions. Hence, we have estimated the asymptotic direction of these particles using a back-trace method \citep{Smart2005} based on IGRF12. The computed directions are shown with colored triangles in Figure~\ref{fig:preissergeo}, where the color code corresponds to the energy. As expected, the low-energy particles are subjected to large deviations inside the geomagnetic field and therefore their asymptotic directions are almost opposite to that of the of the particles with median energy. 

\begin{figure}[ht!]
\centering
\includegraphics[width=0.8\columnwidth]{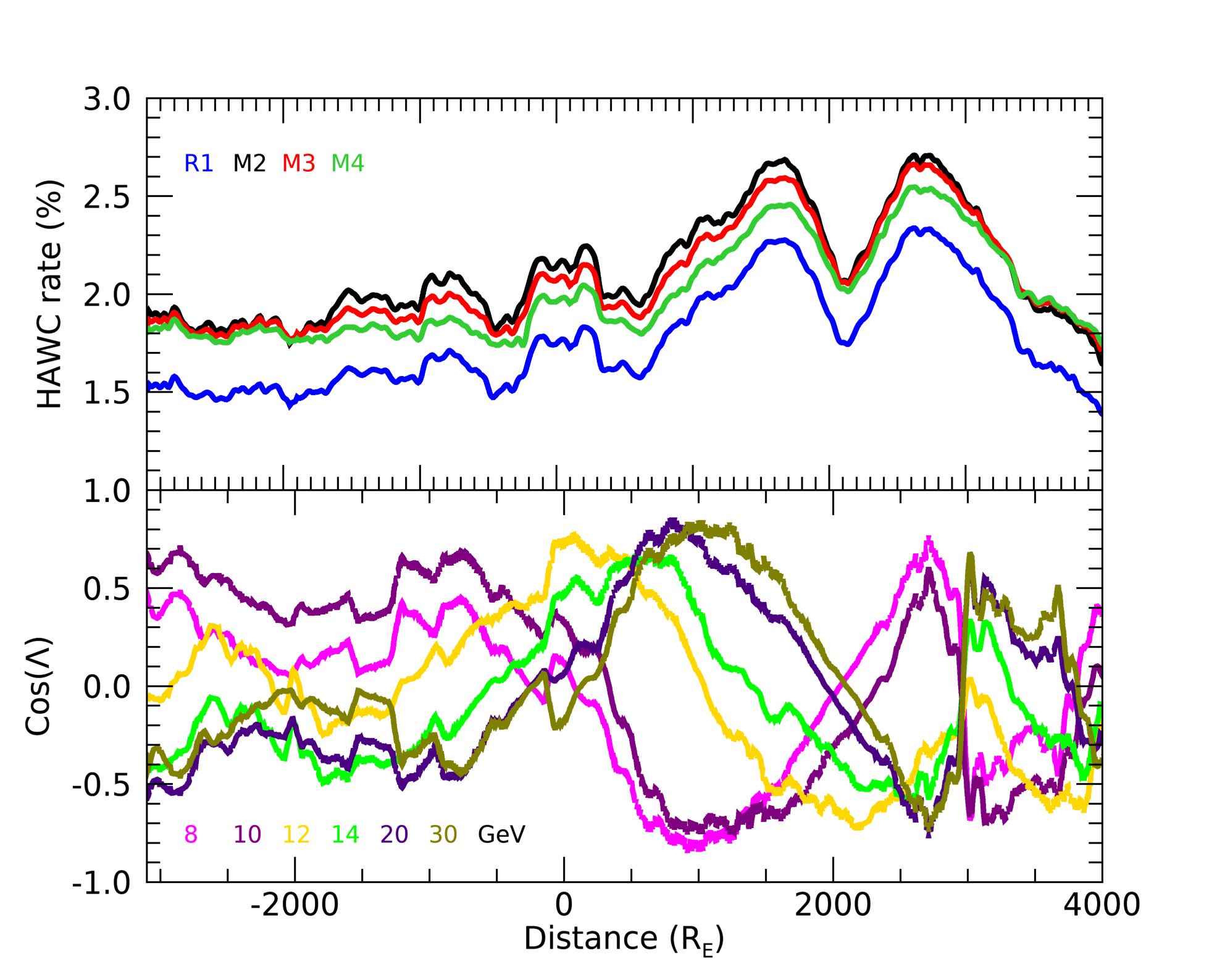}
\caption{The top panel shows the percentage deviation of the HAWC TDC-scaler rates: R1 (blue), Multiplicities 2, 3 and 4 (black, red and green, respectively) during the passage of the MFR at the Earth vicinity. Whereas the bottom panel shows the cosine of the alignment angle ($\Lambda$) between the HAWC asymptotic direction and the interplanetary $B-$field 
during this Earth-MFR encounter, the particle energies are marked with different colors. The distance was computed taking in to account the velocity of the MFR.
\label{fig:aln}}
\end{figure}

The alignment angle ($\Lambda$) between the normal vector of the asymptotic direction ($\hat{N}$) and the interplanetary $B-$field  can be represented as 
\begin{equation}
  \cos (\Lambda) ~ = ~ \frac{\hat{N}\cdot\vec{B}}{|B|}.
\end{equation}

The cosine of $\Lambda$ reaches the value 1 when $\vec{B}$ is parallel to the HAWC asymptotic direction, allowing the detection of particles of the specific energy (marked with colors in Figure~\ref{fig:aln}). On the other hand, when $\Lambda < 1$ the particles are not able to reach the detector. It is important to note that during the local minimum at the middle of the double peak structure, $\cos(\Lambda) \leq 0$ and the parallel distance traveled was $\le  2000 R_E$ (Figure \ref{fig:aln}) for all particle energies. At this point, the MFR axis was perpendicular to the observing directions of HAWC. Considering this alignment, the first peak of the double peak structure was dominated by particles with energy in the range $\sim$ 14-30 GeV range, whereas the second peak was dominated by lower-energy particles. From this analysis (and as seen in Figure~\ref{fig:aln}) we conclude that the double peak structure observed by HAWC is a geometric effect due to the alignment between the MFR axis and the HAWC asymptotic direction. 

\section{Discussion}
\label{sec:discussion}

Assuming the intrinsic GCR population is stationary,the local interstellar spectrum of low-rigidity (tens of GV) GCRs is isotropic and constant. Therefore, any change in the intensity at the top of Earth’s atmosphere is due to solar modulation. At ground level, the measurements of the GCR spectrum are made using secondary particles, therefore the atmospheric conditions must be taken into account. In this way, once the signal is corrected for atmospheric effects, only one source of GCR fluctuations remains: the Sun.

It is well known that the presence of the largest transients in the SW can cause decreases in the GCR intensity. Although there has been controversy about how the ICME components (shock, sheath, and MFR/CME) are related to this flux modulation \citep{Cane2000}, nowadays it is clear that the  decreases in GCRs observed at high energy ($\gtrsim$ 10 GeV) are mainly attributed to MFRs/ejections and the cumulative diffusion mechanism \citep{Arun2013}, and at lower energies the decreases are caused by the shielding mechanism of the shock-sheath system \citep{Wibberenz1998}.

However, we  have described here an interesting phenomenon that has received little attention. This is a local enhancement of the GCR intensity (as opposed to an FD)   associated  with the passage of an ICME over the Earth.

Such GCR enhancements associated with both FDs and geomagnetic storms were observed and systematically studied \citep[e.g.][]{Kondo1960}, and at that time the thought was that they were a result of changes in cutoff rigidity due to variations of the Earth's magnetic field during geomagnetic storms. This view was eventually discarded because polar stations observed similar enhancements \citep{Dorman1963}.

During the 2016 October  event, 
as shown by the satellite data (Section \ref{sec:magnetosphere}) and D$_{ST}$ index (bottom panel of Figure \ref{fig:fluxropeSW}) 
the major geomagnetic disturbances 
occurred the day before  the HAWC observation, confirming that the GCR enhancement was not caused by the geomagnetic storm or changes in the cutoff rigidity,
although, we note that the geomagnetic field was still disturbed during the event but by a much smaller amount.

We propose that the GCR enhancement observed by HAWC was due to the anisotropic distribution of GCR particles in the MFR. We believe that the GCRs entered the MFR isotropically (taking into account that the shielding of the shock/sheath region is negligible) and were then guided along the helical geometry of the force-free field  for a considerable distance before they escaped (see Figure~\ref{fig:cartoon}).
 An anisotropic GCR flux  was found during the FD observed in 2013 April. This increase in GCRs detected by the neutron monitor network was  analyzed by \citet{2018ApJ...852L..26T}, these authors attributed the observed anisotropy to guiding center drifts of energetic particles inside an MFR, as predicted by \citet{2009ApJ...704..831K}.

 \citet{Belov2015}  studied the effects of 99 MCs on the density and anisotropy of 10 GV CRs observed during solar cycle 22 and 23.  They found a general increase of the mean anisotropy (2.03\% versus 1.38\% for MC and non-MC ICMEs, respectively), and more important they found that 20\% of the events showed an increase in the CR density. Even though they did not find an explicit relationship between the maximum magnetic field and the CR density extrema, they established that the events with a clear increase of the CR density have a maximum magnetic field $<$ 18 nT. We note that  the mean value of the magnetic field during the 2016 October event was $\sim$ 18 nT, and it  reached a maximum value of $\sim$~24 nT. The difference may be due to the fact that this event took place during the declining phase of solar cycle 24 which was more weaker than the previous cycles, and therefore the heliospheric conditions were different.
From a theoretical point of view \citet[][and references therein]{doi:10.1029/2018JA025964} developed a model to reproduce the decrease in the local GCR population inside an MFR and found large anisotropies that depend on the MFR geometry and magnetic field strength. Then, to determine the GCR anisotropies during the passage of the MFR observed in 2016 October  \citep[see][for a discussion of the anisotropies during FDs]{Lockwood1971}, we constructed a model to track the trajectories of GCRs inside this specific MFR .
 
To support the idea of the anisotropic distribution of GCR particles in the MFR, we first recall the special circumstances that gave rise to the HAWC detection.

As discussed in Section~\ref{sec:sun}, a quiet filament eruption was the solar origin of the observed MFR. The MFR structure of the associated CME is evident in the SECCHI image in the panel (f) of Figure~\ref{fig:sun}.  Then, as shown in Section~\ref{sec:sim}, the transport of this MFR in the interplanetary medium was surrounded, without interaction, by two high speed streamers, helping in this way to maintain an helical and strong magnetic field, low density, low turbulence and a large magnetic/flow pressure ratio. Next, in Section \ref{sec:magnetosphere} we have characterized the response  of the $B-$field in the Earth's magnetosphere due to its interaction with the MFR, and shown that even though the magnetosphere was disturbed during the GCR enhancement, the major disturbances and the geomagnetic storm were observed the previous day, confirming in this way that the observed GCR enhancements are not related to changes of the cutoff rigidity due to the geomagnetic storms. 

These facts, and most importantly the alignment between the MFR axis and the asymptotic direction of the HAWC site, allowed the charged particles that were redirected and guided by the MFR (Section~\ref{sec:model}) to reach deep into the low Earth atmosphere to be detected by sensitive ground-level detectors such as HAWC.

\section{Conclusions}
\label{sec:conclusion}

On 2016 October 14 the TDC-scaler system of HAWC registered an unusual increase of the GCR flux. In this work, we have presented evidence that the observed enhancement was caused by an anisotropy generated inside an interplanetary MFR, by the guiding of the GCRs along the axis of the large-scale magnetic structure.

This detection was made possible by a set of unusual circumstances:
\begin{itemize}
\item
The quiet filament eruption gave rise to the slow CME that reached the Earth with a weak discontinuity (shock) and a relatively quiet sheath region.

\item
The interaction-free transport of the MFR between two high-speed streamers prevented its deformation because the streamer in front swept the structures of the ambient SW while the one behind   did not allow any other faster structure to reach this slow ICME.

\item
The magnetic field configuration and 
the low turbulence level allowed the long-distance guiding of the GCR inside the MFR.

\item
The alignment between anisotropic GCR flux, parallel to the MFR axis, and the asymptotic direction of the HAWC detector. 

\item 
The high sensitivity achieved by the HAWC TDC-scaler system, which allows the detection of GCR variations with a statistical error $< 0.01 \%$.

\end{itemize}

These conditions allowed HAWC to detect the anisotropic flux of GCR created by the toroidal magnetic field of the MFR. 

\section{Acknowledgments}

We acknowledge use of NASA/GSFC's Space Physics Data Facility's OMNIWeb (or CDAWeb or ftp) service, and OMNI data.

To SOHO, project of international cooperation between ESA and NASA.

The D$_{ST}$ (Disturbance Storm-Time) index used in this paper was provided by the WDC for Geomagnetism, Kyoto (\url{http://wdc.kugi.kyoto-u.ac.jp/wdc/Sec3.html}).

We acknowledge the support from: the US National Science Foundation (NSF); the US Department of Energy Office of High-Energy Physics; the Laboratory Directed Research and Development (LDRD) program of Los Alamos National Laboratory; L.P. thanks CONACyT grants: 174700, 271051, 232656, 260378, 179588, 254964, 258865, 243290, 132197, A1-S-46288, A1-S-22784, c{\'a}tedras 873, 1563, 341, 323, Red HAWC, M{\'e}xico; DGAPA-UNAM grants AG100317, IN111315, IN111716-3, IN111419, IA102019, IN112218; VIEP-BUAP; PIFI 2012, 2013, PROFOCIE 2014, 2015; the University of Wisconsin Alumni Research Foundation; the Institute of Geophysics, Planetary Physics, and Signatures at Los Alamos National Laboratory; Polish Science Centre grant DEC-2018/31/B/ST9/01069, DEC-2017/27/B/ST9/02272; Coordinaci{\'o}n de la Investigaci{\'o}n Cient{\'i}fica de la Universidad Michoacana; Royal Society - Newton Advanced Fellowship 180385. Thanks to Scott Delay, Luciano D{\'i}az, Eduardo Murrieta and Elsa Noemi S{\'a}nchez for technical support.

\bibliography{sample63}

\appendix

\section{Magnetospheric signatures of the Flux Rope} 
\label{app:magnetosphere}

Although the October 2016 event was a slow ICME, it caused a large and unusual Earth magnetospheric disturbance as seen by five spacecraft located inside and outside the magnetosphere. Here we present a detailed analysis of these observations.

The arrival of the shock produces a step-like increment in the Earth's magnetic field due to the increment in $P_{FLOW}$ of the former between the CME driven shock and through the sheath region of the CME (red curve in the fourth panel of Figure~\ref{fig:fluxropeSW}) which compress the Earth's magnetic field. This signature can be observed on the first panel of Figures~\ref{fig:preissergo15},~\ref{fig:preisserthe_e}~and~\ref{fig:preissermms1} (marked as a gray vertical line) corresponding to GOES-15, THEMIS-E and MMS-1 which were located inside the day-side magnetosphere. THEMIS-C located in the magnetosheath night-side region also observed the arrival of the shock and the whole structure of the CME (see Figure~\ref{fig:preisserthe_c}) since the Earth's magnetic field did not permeate that region as can be corroborated from the upstream/downstream values for different spacecraft (locations) in Table~\ref{tabla:spcrft}. 

The detection times of this discontinuity as reported also in Table~\ref{tabla:spcrft} are in agreement with the relative locations of all the spacecraft as well as with the 9$^\circ$ angle of the major axis of the MFR with respect to the Y-axis (see Section \ref{subsec:ICME}) as can be observed in the panel corresponding to XY-plane of Figure~\ref{fig:preisserorbits}.

The data are in GSE-coordinates except for GOES-15 which is presented in E (earthward), P (parallel) and N (normal) coordinate system, where N (our like-$B_Z$ component) points northward, perpendicular to the spacecraft orbit plane. 

\begin{figure}[htpb] 
\includegraphics[width=0.7\columnwidth]{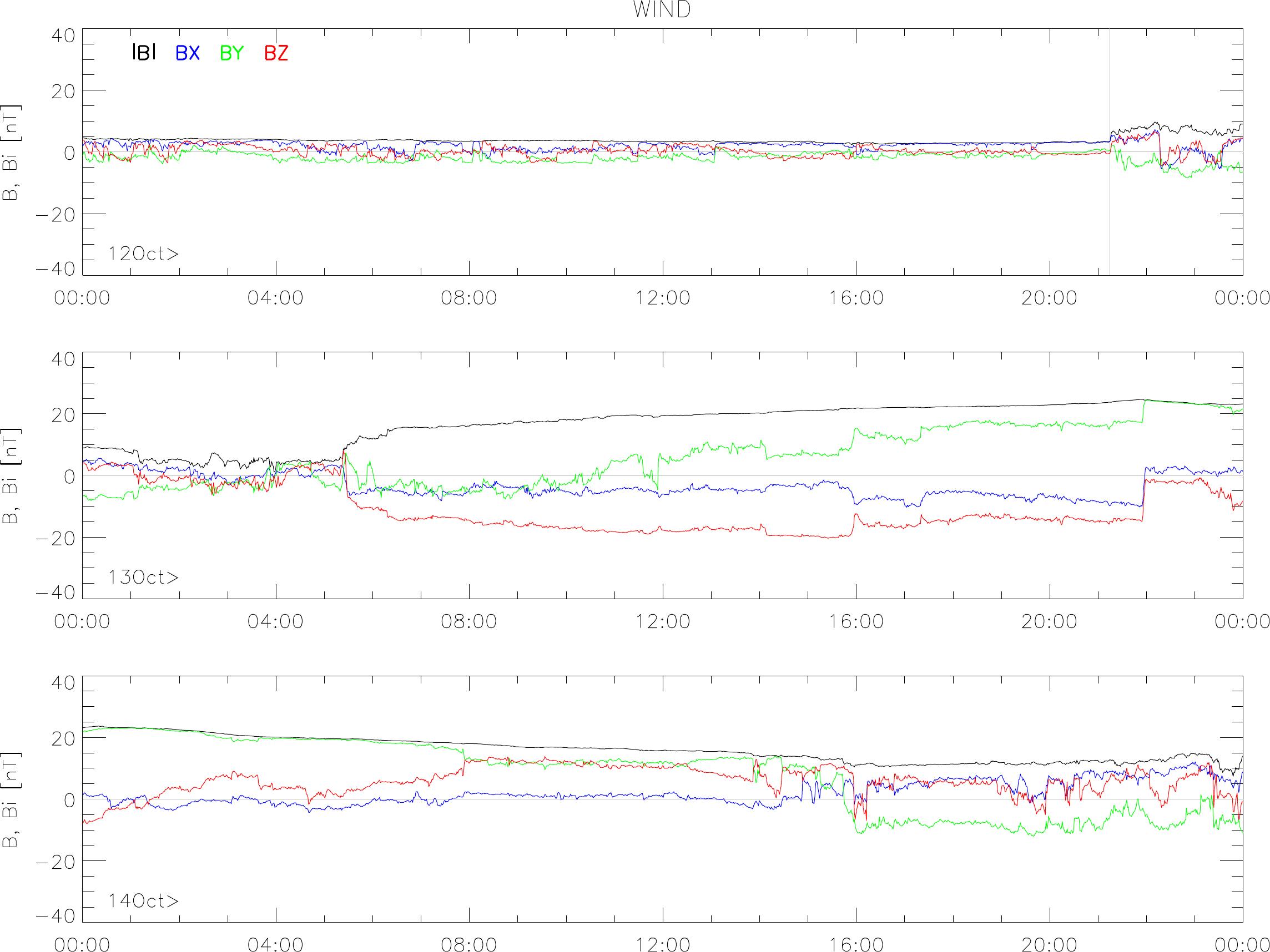} 
\centering
\caption{Magnetic field magnitude (black) and its components (colors) in GSE coordinates observed by Wind spacecraft on October 12 (top), 13 (middle) and 14 (bottom). The vertical gray line indicates the forward shock at 21:15:37 UT on October 12,  2016. 
\label{fig:preisserwind}}
\end{figure}

\begin{figure}[htbp]
\includegraphics[width=0.7\columnwidth]{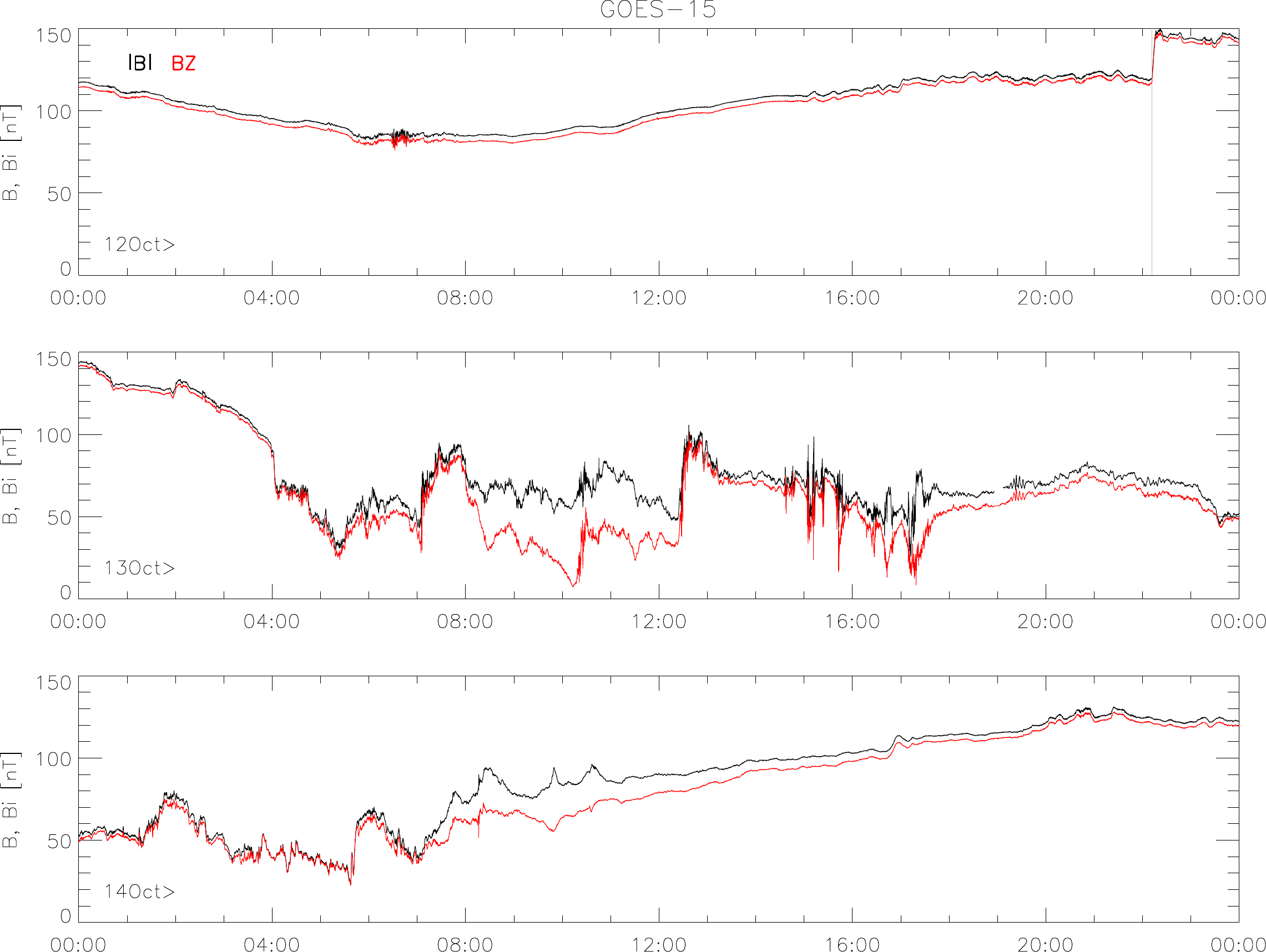} 
\centering
\caption{Magnetic field magnitude (black) and $B_Z$ component (red) in EPN coordinates observed by magnetometer-1 on GOES-15 spacecraft on October 12 (top), 13 (middle) and 14 (bottom). The vertical gray line indicates the CME-shock related discontinuity at 22:11:51 UT on October 12, 2016.  
\label{fig:preissergo15}}
\end{figure}

\begin{figure}[htbp]
\includegraphics[width=0.7\columnwidth]{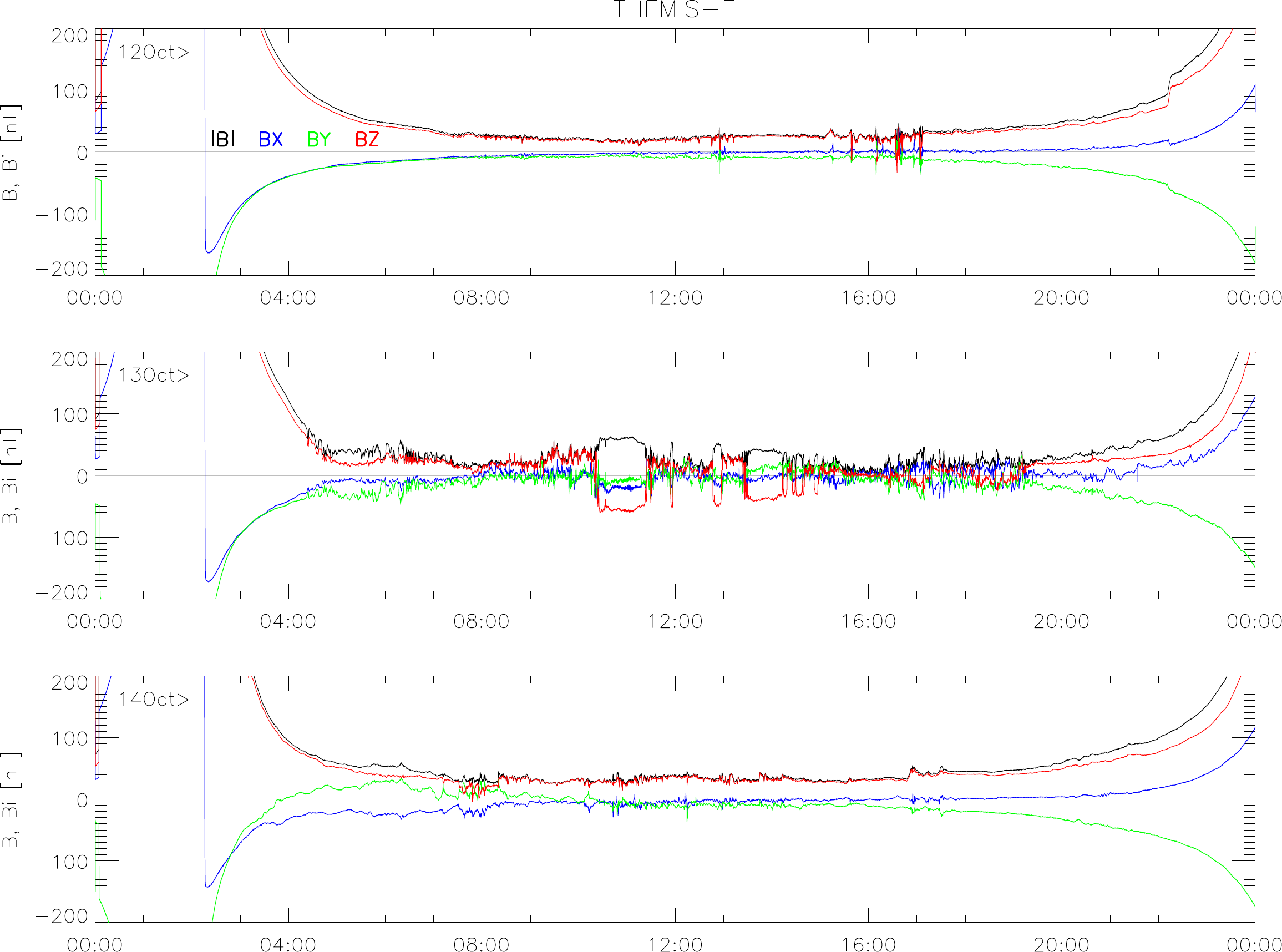}
\centering
\caption{Magnetic field magnitude (black) and its components (colors) in GSE coordinates observed by THEMIS-E spacecraft on October 12 (top), 13 (middle) and 14 (bottom). The vertical gray line indicates the CME-shock related discontinuity at 22:13:33 UT on October 12,  2016.  
\label{fig:preisserthe_e}}
\end{figure}

\begin{figure}[htbp]
\includegraphics[width=0.7\columnwidth]{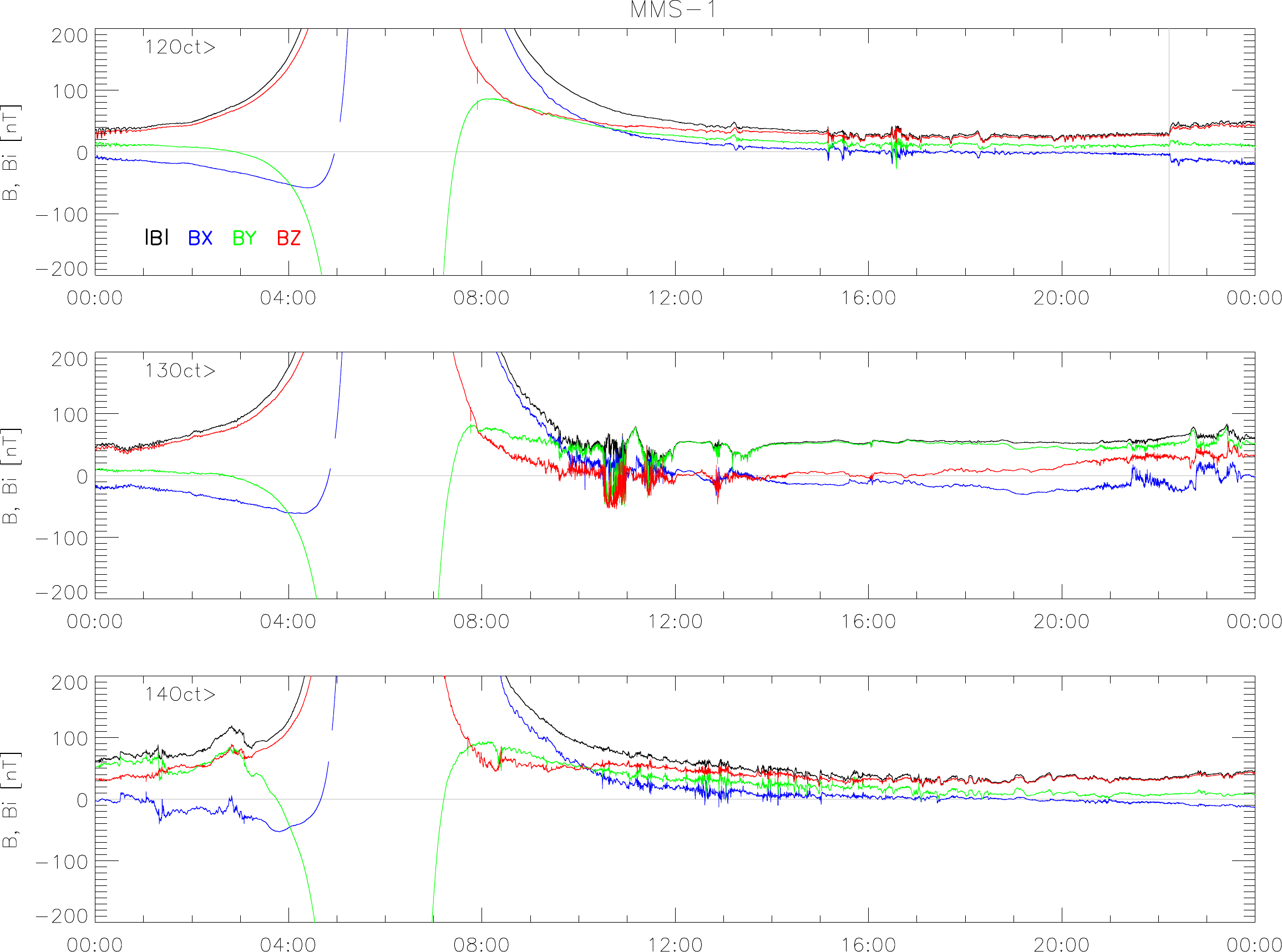} 
\centering
\caption{Magnetic field magnitude (black) and its components (colors) in GSE coordinates observed by MMS-1 spacecraft on October 12 (top), 13 (middle) and 14 (bottom). The vertical gray line indicates the CME-shock related discontinuity at 22:13:52 UT on October 12,  2016.
\label{fig:preissermms1}}
\end{figure}

\begin{figure}[htbp]
\includegraphics[width=0.7\columnwidth]{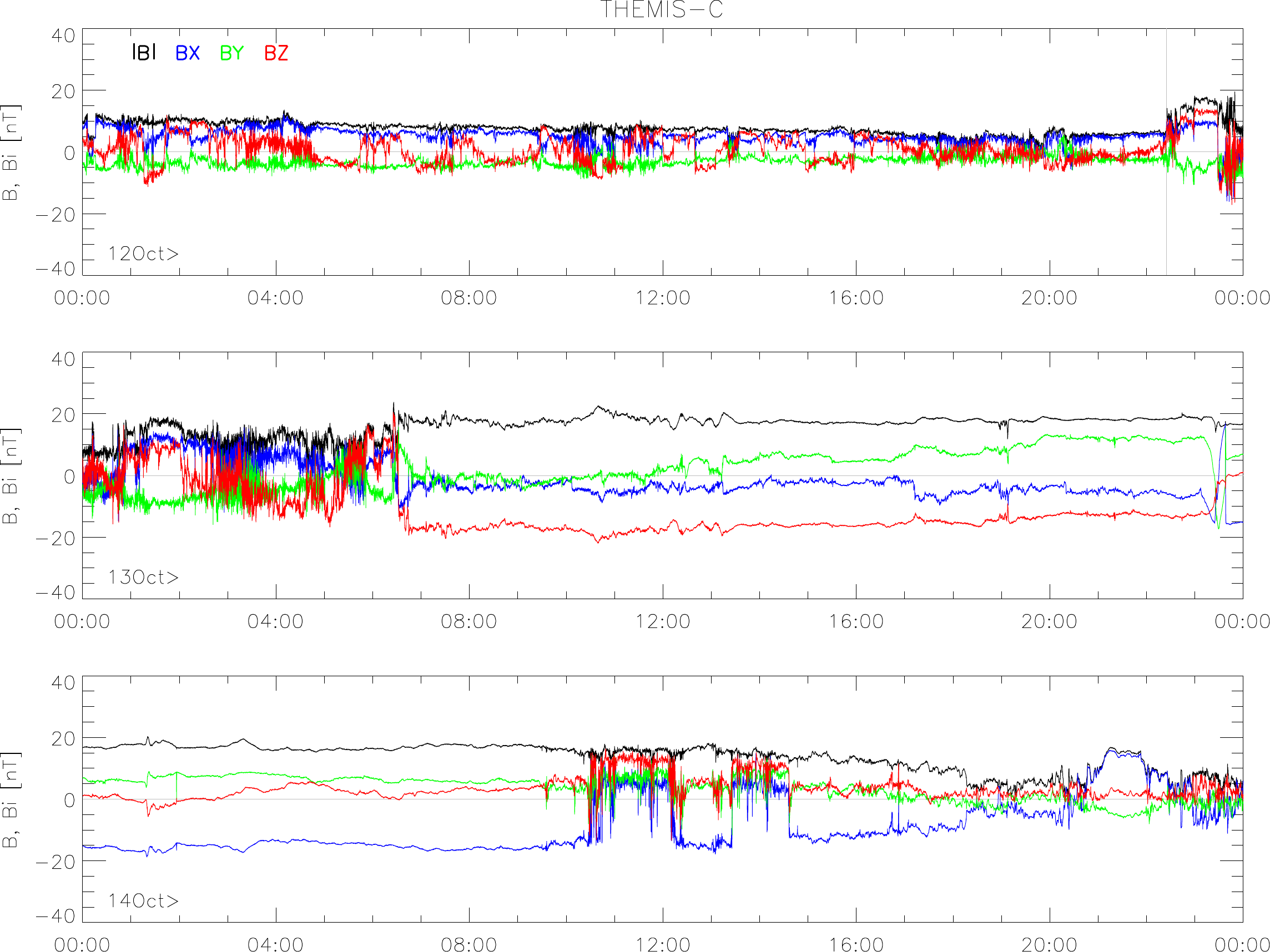} 
\centering
\caption{Magnetic field magnitude (black) and its components (colors) in GSE coordinates observed by THEMIS-C spacecraft on October 12 (top), 13 (middle) and 14 (bottom). The vertical gray line indicates the CME-shock related discontinuity at 22:24:49 UT on October 12,  2016.  
\label{fig:preisserthe_c} }
\end{figure}

\newpage
\section{Estimating axial distance} 
\label{app:axial}
 
\begin{figure}[htbp]
\includegraphics[width=0.8\columnwidth]{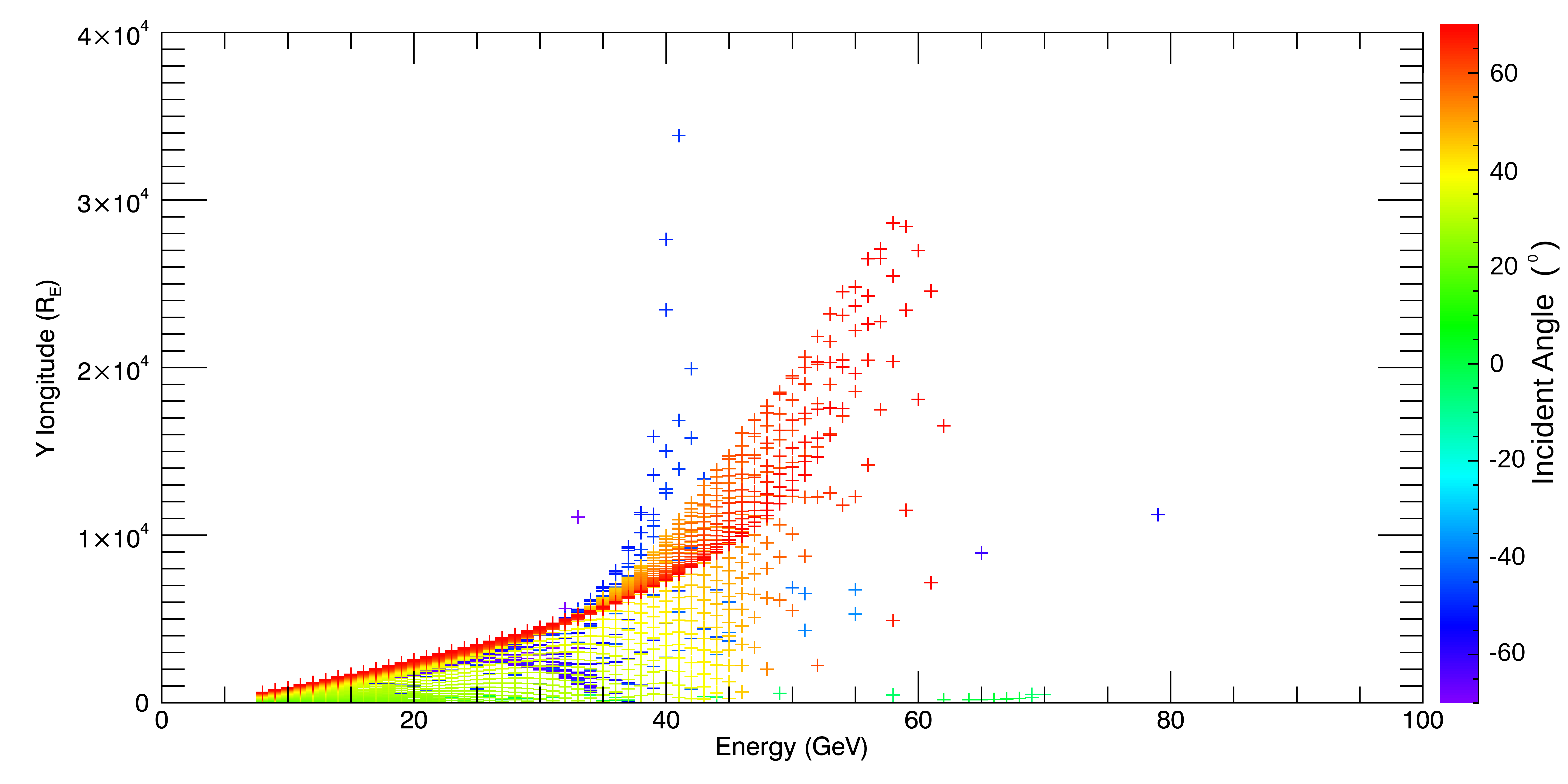}
\centering
\caption{The axial distance traveled by a particle as they enter the MFR at point, $x= -0.5 \times S_{MFR}$, $x=0$ and $z=0$. The color code corresponds to the incident angle in degrees between X and Y, at which the particles entered the MFR. \label{EnY}}
\end{figure}

The GCR trajectory simulations were performed for protons of energy ranging from 8-100 GeV. Protons were thrown in to the MFR with initial position at $x= -0.5 \times S_{MFR}$, $y=0$ and $z=0$. A wide range of incident angles ($\Psi$) in the XY-plane  are considered from -$70^{\circ}$ to $70^{\circ}$ with an increment of $1^{\circ}$. We checked the alignment (inside the $85^{\circ} < \Psi < 95^{\circ}$ cone) of particle trajectory along the axis of MFR, and estimated the distance each particle traveled along the axial direction. The distance traveled along the axial direction for different energy particle are shown in Figure~\ref{EnY}. From this we can conclude that particle of energies $>$ 60 GeV are less likely to get aligned with the magnetic topology of the MFR we are considering.  The increments in the TDC-scaler rates observed are mainly contributed by lower energy particles ($\leq 60$ GeV). 
 
\end{document}